\documentclass[pra, aps, twocolumn, preprintnumbers, superscriptaddress, showkeys, 10pt, nofootinbib]{revtex4-2}

\usepackage[dvipsnames]{xcolor}
\definecolor{mediumpersianblue}{rgb}{0.0, 0.4, 0.65}
\definecolor{persianred}{rgb}{0.8, 0.2, 0.2}
\usepackage[utf8]{inputenc}
\usepackage{textcomp}

\usepackage[colorlinks=true,citecolor=mediumpersianblue,linkcolor=persianred, urlcolor=persianred]{hyperref}

\usepackage{graphicx}
\usepackage{hyperref}
\usepackage{multirow}
\usepackage{amsmath,amsfonts}
\usepackage{mathrsfs}
\usepackage{textcomp}
\usepackage{algorithm2e}
\usepackage{booktabs}
\usepackage{algpseudocode}
\usepackage{listings}
\usepackage{natbib}
\usepackage{physics}
\usepackage{tikz}
\usetikzlibrary{quantikz2}
\tikzstyle{block} = [rectangle, rounded corners, minimum width=3cm, minimum height=1cm, text centered, draw=black]
\tikzstyle{arrow} = [thick,->,>=stealth]

\begin{document}

\title{Quantum noise modeling through Reinforcement Learning}

\newcommand{\SUR}{Dep. of Physics, La Sapienza University of Rome, Piazzale Aldo Moro 2, Rome, 00185, Italy}

\newcommand{\TII}{Quantum Research Centre, Technology Innovation Institute, Abu Dhabi, UAE.}

\newcommand{\ANU}{School of Computing, The Australian National University, Canberra, ACT, Australia}

\newcommand{\IFT}{Instituto de F\'isica Te\'orica, UAM-CSIC, Universidad Aut\'onoma de Madrid, Cantoblanco, Madrid, Spain}

\newcommand{\INFN}{Sez. di Roma, INFN, Piazzale Aldo Moro 2, Rome, 00185, Italy}

\newcommand{\CERNaff}{CERN, Theoretical Physics Department, CH-1211 Geneva 23, Switzerland.}

\newcommand{\MIaff}{TIF Lab, Dipartimento di Fisica, Universit\`a degli Studi di Milano, Italy}

\newcommand{\INFNUNIMI}{INFN, Sezione di Milano, I-20133 Milan, Italy.}

\author{Simone Bordoni}
\thanks{These authors contributed equally to this work.}
\affiliation{\SUR}
\affiliation{\TII}
\affiliation{\INFN}

\author{Andrea Papaluca}
\thanks{These authors contributed equally to this work.}
\affiliation{\ANU}
\affiliation{\TII}

\author{Piergiorgio Buttarini}
\thanks{These authors contributed equally to this work.}
\affiliation{\SUR}
\affiliation{\TII}

\author{Alejandro Sopena}
\thanks{These authors contributed equally to this work.}
\affiliation{\TII}
\affiliation{\IFT}

\author{Stefano Giagu}
\affiliation{\SUR}
\affiliation{\INFN}

\author{Stefano Carrazza}
\thanks{Corresponding author: stefano.carrazza@unimi.it}
\affiliation{\CERNaff}
\affiliation{\MIaff}
\affiliation{\INFNUNIMI}
\affiliation{\TII}

\preprint{TIF-UNIMI-2024-9}

\begin{abstract}
  In the current era of quantum computing, robust and efficient tools are essential to bridge the gap between simulations and quantum hardware execution. In this work, we introduce a machine learning approach to characterize the noise impacting a quantum chip and emulate it during simulations. Our algorithm leverages reinforcement learning, offering increased flexibility in reproducing various noise models 
  compared to conventional techniques such as randomized benchmarking or heuristic noise models. 
  The effectiveness of the RL agent has been validated through simulations and testing on real superconducting qubits. Additionally, we provide practical use-case examples for the study of renowned quantum algorithms.
\end{abstract}

\keywords{Machine Learning; Reinforcement Learning; Quantum Computing; Quantum Noise}

\maketitle

\section{Introduction}
One important unsolved technological question regards the practical applicability of Noisy Intermediate Scale Quantum~(NISQ)~\cite{Preskill_2018} computers. 
Despite the expectation that quantum computers will outperform classical computers in certain computational 
tasks~\cite{Shor_1997, grover1996fast, Zhong_2020}, the usability and reliability of NISQ devices are hindered by a large error rate.
These errors arise from gate infidelities, unwanted environmental interactions, thermal relaxation, measurement errors, and cross-talk~\cite{PhysRevLett.121.090502, doi:10.1146/annurev-conmatphys-031119-050605, PRXQuantum.3.020301, PhysRevB.104.045420, 9355264}.
Currently, there are techniques to mitigate these errors~\cite{PRXQuantum.2.040326, PhysRevA.105.062404, 10.1145/3352460.3358265, Sopena_2021, PRXQuantum.2.040330, PhysRevA.106.012423}. 
However, it has been demonstrated that any quantum circuit for which error mitigation is efficient must be classically simulable~\cite{schuster2024polynomialtimeclassicalalgorithmnoisy}.
Therefore, it is widely regarded that quantum advantage will only be achieved with future generations of fault-tolerant quantum devices~\cite{article_knill, PhysRevX.8.021054, PhysRevA.86.032324, Varsamopoulos_2017, bordoni2023, cacioppo2023quantumdiffusionmodels}.
Despite the imperfect results obtained on NISQ devices, numerous algorithms have been developed and implemented on this hardware. Machine learning-inspired models, in particular, have shown promising results in recent years~\cite{Biamonte_2017, 8939749, Cerezo_2021, egger2023study, particles6010016, Robbiati:2853183, robbiati2022quantum, cruzmartinez2023multivariable}.\\
Testing and developing new algorithms in the NISQ era can be challenging due to limited access to quantum chips. Furthermore, the currently accessible quantum computers in the cloud are highly demanded, resulting in lengthy waiting queues~\cite{9668289}. In this context, emulating the noise of these devices emerges as an alternative to accelerate circuit testing.

This work aims to develop a model capable of learning hardware-specific noise for use in circuit simulations. Our approach employs Reinforcement Learning~(RL)~\cite{sutton2018reinforcement, kaelbling1996reinforcement, wiering2012reinforcement} to train an agent to add noise channels that replicate the noise pattern of a specific quantum chip. This method minimizes heuristic assumptions about the noise model, thereby enhancing the adaptability and generalization properties.
The algorithm has been validated on both simulations and real quantum devices hosted at the Quantum Research Center~\cite{qibolab_platforms_qrc} of the Technology Innovation Institute~(TII) in Abu Dhabi, demonstrating its ability to accurately predict different noise patterns. 

The outline is as follows. 
Section~\ref{sec_background} introduces the basic concepts necessary to understand the proposed algorithm, noise in quantum circuits and reinforcement learning. Moreover, we review previous works on similar topics to provide more context of the work in relation to recent applications of reinforcement learning in quantum computing and noise simulation techniques.
Section~\ref{sec_methodology} provides a detailed description of the reinforcement learning algorithm, focusing on its training and use for noise prediction.
Section~\ref{sec_results} presents the results obtained with the proposed algorithm on both simulations and real quantum devices, and compares its performance with other noise predictors.
Section~\ref{sec_applications} demonstrates examples of the algorithm's use-cases for famous quantum algorithms.

The Qibo framework~\cite{Efthymiou_2021, Efthymiou_2022} was used for the realization of this work. 
It provides \texttt{Qibo}~\cite{Efthymiou_2021, Efthymiou_2022} as a high-level language API for crafting quantum computing algorithms, 
\texttt{Qibolab}~\cite{efthymiou2023qibolab,pedicillo2024} as a tool for quantum control, 
and \texttt{Qibocal}~\cite{pasquale2023opensource} for conducting quantum characterization and calibration routines.
\\
All the code developed for this work is available on GitHub~\footnote{\url{https://github.com/qiboteam/rl-noisemodel}}. It is possible to use the code to reproduce the results as well as testing the algorithm under different noise conditions. The code is intended to be easily customizable in order to allow users to define their own RL agents for specific applications.

\section{Background}\label{sec_background}
This section provides an overview of the essential concepts required to understand the proposed noise simulation algorithm. Specifically, we briefly discuss noise in quantum circuits, the randomized benchmarking technique and introduce reinforcement learning.
We also review some related results obtained in the context of noise simulation and reinforcement learning applied to quantum computing.

\subsection{Related works}\label{sec_previous}
The characterization and modeling of quantum noise is an important research area that has yielded significant results in recent years \cite{inproceedings_caracterization, PhysRevA.104.062432, Harper_2020}.
While standard noise tomography techniques such as randomized benchmarking are useful for obtaining a standardized evaluation of noise, they lack the ability to capture specific characteristics that are essential for realistic simulations.
These techniques can be improved by incorporating non-Markovian noise and realistic finite-time gate models, as proposed by Brillant et al.~\cite{brillant2025randomizedbenchmarkingnonmarkoviannoise}. Other studies focus on reducing the resources required to perform these techniques and on maximizing the information extracted from experimental data \cite{Onorati_2023, huang2025randompulsesequencesqubit, Chen_2023}.\\
Another line of research involves developing efficient algorithms to fit a Lindbladian to tomography data \cite{malekakhlagh2025efficientlindbladsynthesisnoise, Onorati_2023, van_den_Berg_2024}, providing compact noise models and quantitative measures of non-Markovianity.
Finally, it is worth mentioning the tensor network approach proposed by Mangini et al. \cite{Mangini_2024}, which enables the study of correlated noise in systems with up to 20 qubits.\\
Various techniques based on machine learning have also been proposed for noise characterization. For instance, Zlokapa et al. introduced a deep learning method to identify noise patterns, aiming to reduce noise levels during the transpilation process \cite{zlokapa2020deep}.
Canonici et al. trained machine learning models on measurement data to infer noise parameters in neutral-atom quantum processors and additionally employed a reinforcement learning agent to design corrective control pulses that mitigate measurement noise \cite{Canonici_2023}.

Separately, reinforcement learning has been explored for a variety of quantum control and calibration tasks. In quantum gate design, several studies apply deep RL to optimize pulse shapes or gate parameters. For instance, Niu et al. trained an RL agent to optimize two-qubit gate pulses in the presence of leakage and stochastic errors \cite{niu2018universalquantumcontroldeep}. In superconducting systems, Baum et al. experimentally employed deep RL to autonomously design single- and two-qubit gates without requiring knowledge of the device Hamiltonian \cite{PRXQuantum.2.040324}.\\
Reinforcement learning has also been applied to improve qubit readout~\cite{phdthesis_hoffer}. Chatterjee et al. demonstrated on superconducting hardware that an RL agent can tailor readout pulses to achieve state-of-the-art fidelity while significantly reducing readout time \cite{PhysRevApplied.23.054057}.
Similar applications of RL methods have been proposed for automated calibration and tuning. For example, Crosta et al. \cite{crosta2024automaticrecalibrationquantumdevices} developed a model-free RL agent that periodically adjusts calibration parameters to maintain optimal performance, without relying on an explicit noise model of the environment. Related frameworks have been introduced by K. Young \cite{2023APS..MARB72005Y} and Nguyen et al. \cite{Nam_Nguyen_2024} to accelerate the calibration of superconducting logic gates.\\
Finally, several works have applied RL to quantum error correction and quantum algorithm design. Nautrup et al. used RL to adapt surface-code error-correction circuits to unknown noise channels \cite{Nautrup_2019}. Olle et al. extended this approach to discover new error correction codes and encoding circuits tailored to a given hardware's gate set, connectivity, and noise model \cite{olle2024simultaneousdiscoveryquantumerror}.

\subsection{Noise in the quantum circuit model}\label{sec_noise}

Current and near-term quantum computers lack fault tolerance, and their usefulness is limited by the presence of noise and errors.
To understand these limitations, we delve into the quantum circuit model, a framework introduced by Deutsch in 1989~\cite{deutsch_quantum_1989}. This model presumes a quantum register composed of near-ideal qubits. Quantum computations are carried out by altering 
this register through a combination of qubit measurements and unitary operations drawn from a universal set of gates, known as native gate set.
Within this context, we distinguish four types of errors.
The first type, state preparation errors, arise during the initialization of the quantum register. These errors result from the need for rapid reset protocols, which involve coupling qubits to other elements such as 
cavities and measurement devices, leading to deviations from the ideal zero state~\cite{reed_fast_2010}. 
The second type of error is due to the limited precision of measurements, which requires their representation as POVMs with inherent uncertainties, thereby preventing unlimited repeated measurements on the same qubit. 
The third type, qubit decoherence, refers to the loss of quantum superposition due to environmental factors. It is typically modeled by relaxation and dephasing times, $T_1$ and $T_2$, for each qubit. 
However, this model can be insufficient when decoherence is correlated, such as when environmental fluctuations or unwanted interactions between the qubits affect multiple qubits similarly.
Lastly, gate imperfections arise from intrinsic errors and control limitations in implementing single-qubit and two-qubit unitaries. These imperfections are measured by the gate fidelity.

It is common to make certain assumptions about the inherent errors. One typical assumption is to distinguish the ideal gates from the errors, considering them as separate processes. 
It is also often assumed that these errors break down into spatially uncorrelated errors that affect the idle qubits, and an average error that is operation-dependent, impacting only the qubits that are being manipulated.
Within this context, it is useful to classify the various types of noise introduced above into two groups: coherent and incoherent. Coherent noise, typically resulting from minor miscalculations in control parameters, 
tends to produce similar errors in successive executions of a quantum circuit, thereby introducing a systematic bias in the output. It is important to note that coherent noise preserves the purity of the state and, once identified, can be corrected~\cite{Cenedese_2023, Kern_2005, Bravyi_2018}.
On the other hand, incoherent noise can be viewed as processes that cause entanglement between the system and its environment.

The errors and imperfect operations are typically represented using the formalism of quantum channels, i.e., trace-preserving completely positive maps of density matrices into density matrices.
In the ideal operation of a quantum computer, a positive map can be just a unitary transformation $\varepsilon(\rho) = U\rho U^\dagger$, where $U$ describes a quantum gate.
Coherent noise is unitary, and we model it using single-qubit rotation gates ($R_x$, $R_y$, $R_z$). For instance, a coherent error could introduce an unintended deviation $\epsilon$ in the $x$ direction during the application of $R_j(\theta)$, 
altering the state $\rho=R_j(\theta)\rho_0R_j(\theta)^\dagger$ to
\begin{equation}
    \operatorname{Coh}_x(\rho) = R_x(\epsilon)\rho R_x(\epsilon)^\dagger \, .
\end{equation}
We model incoherent noise as local depolarization and amplitude damping.
Depolarization noise tends to drive the state towards the maximally mixed state,
\begin{equation}
    \operatorname{Dep}(\rho)=(1-\lambda)\rho+\lambda\frac{\operatorname{Tr}(\rho)}{2^n}\mathbb{I} \, ,
\end{equation}
where $n$ is the number of qubits and $\lambda$ is the depolarization parameter.
Amplitude damping noise models the loss of energy from the qubit to the environment, and it is described by the map
\begin{equation}
    \operatorname{Damp}(\rho) = A_1\rho A_1^\dagger + A_2\rho A_2^\dagger \, ,  
\end{equation}
where $A_1 = \ket{0}\bra{0} + \sqrt{1-\gamma}\ket{1}\bra{1}$ and $A_2 = \sqrt{\gamma}\ket{0}\bra{1}$, and $\gamma$ represents the decay probability from $\ket{1}$ to $\ket{0}$.
While it is possible to add more complexity to the modeling of incoherent noise, we aim for an effective description with few parameters to prevent overfitting.
In particular, preliminary tests showed no significant performance enhancement when adding the thermal relaxation channel or other channels that depend on multiple parameters. Moreover, the high number of parameters introduced by this kind of channels make the model more prone to overfitting the noise model and increases the action space of the agent thus complicating the training process.

Multiple noise channels can be combined in order to construct complex noise models~\cite{PhysRevA.104.062432, article_model_realistic}. 
The parameters of these channels are usually inferred from calibration results.
In this work, we show that it is possible to use ML to fit a noise model directly from the execution of quantum circuits.

\subsection{Randomized benchmarking}\label{sec_rb}

Randomized benchmarking (RB) is widely recognized as one of the most effective experimental techniques for evaluating the quality of quantum operations in quantum computing platforms~\cite{Emerson_2005, 2019npj, helsen_general_2022, heinrich2023randomized}. 
Its widespread adoption is due to practical advantages, particularly its polynomial scalability with the number of qubits and its robustness to state preparation and measurement errors.

At the core of the RB protocol lies a gateset $G$, typically chosen to be a finite group of quantum gates such as the Clifford group~\cite{magesan_characterizing_2012}. 
The primary objective of RB is to estimate the average gate fidelity of this gateset, providing a reliable measure of the performance of the underlying quantum device.

The protocol proceeds by randomly sampling a sequence of $m$ gates from the gateset $G$, applying them to an initial quantum state $\rho$, and then applying an additional gate designed to invert the net effect of the sequence in the absence of noise. 
This last gate ensures that the final state should ideally return to $\rho$ if all gates were implemented perfectly. 
The actual final state is then measured to estimate the survival probability $p_m$, defined as the probability of finding the system in the initial state after the noisy evolution.
This process is repeated over many random sequences of the same length $m$, and the resulting probabilities are averaged to obtain a single estimate of $p_m$. 
By varying the sequence length $m$, one obtains a decay curve, which reflects the accumulation of errors and allows extraction of the average fidelity of the gateset.

The Clifford group is typically chosen as the gateset $G$ in RB because it enables a simple model for how errors accumulate. 
Under the assumption of gate-independent, completely positive and trace-preserving (CPTP) noise, the average survival probabilities $\{p_m\}$ can be well-approximated by a single exponential decay of the form 
\begin{equation}
    p_m = a f^m + b \, ,
\label{eq:rb_decay}
\end{equation}
where $a$ and $b$ depend on state preparation and measurement errors, and the decay parameter $f$ reflects the average fidelity of the implemented gates~\cite{magesan_characterizing_2012}. 
This behavior arises because averaging over the Clifford group effectively transforms arbitrary noise into a depolarizing channel, which accumulates in an exponential decay with the sequence length $m$. 
This property, unique to unitary 2-designs like the Clifford group, makes it especially well-suited for benchmarking purposes~\footnote{In RB protocols that employ non-Clifford gatesets, the standard exponential decay model~\eqref{eq:rb_decay} no longer holds. Instead, the decay behavior must be described by a generalized multi-exponential form $p_m=\sum_{\lambda\in J_G}A_\lambda f_\lambda^m$, where $J_G$ is an index set determined by the structure of the gateset, the $f_\lambda$ are gateset-dependent quality parameters, and the coefficients $A_\lambda$ depend only on measurement errors~\cite{helsen_general_2022}.}.

Randomized benchmarking can be repurposed as a simplified noise modeling tool by treating the average error channel as an effective depolarizing channel with depolarizing rate $1-f$. 
In this approach, a depolarizing channel with depolarizing rate $1-f$ is inserted after each gate to emulate the observed error behavior (see Appendix~\ref{app:rb}). 
We adopt this technique to construct a baseline depolarizing noise model, which serves as a reference for evaluating our algorithm. 
While this model simplifies the noise landscape by projecting all errors onto a single depolarizing channel, potentially overlooking coherent or gate-dependent contributions, it remains a valuable benchmark for comparison. 
Importantly, we do not sample gates randomly from the Clifford group; instead, we work with a fixed set of Clifford gates native to the quantum hardware under study. 
In this setting, the assumptions underlying standard RB theory may not hold, and the noise may not fully depolarize. 
Consequently, the RB decay may deviate from a pure exponential form. 

\subsection{Reinforcement learning}\label{sec_rl}

Reinforcement learning is a powerful Machine Learning~(ML) paradigm that trains an agent to make optimal decisions in a dynamic environment.
From a mathematical perspective, a reinforcement learning algorithm is described as a Markov Decision Process~(MDP)~\cite{PUTERMAN1990331}. 
An MDP is characterized by a tuple $(S, A, P, R, \gamma)$, where $S$ represents the set of possible environmental states, $A$ denotes the set of actions, 
$P(s^\prime |s,a)$ is the transition probability of reaching state $s^\prime$ from state $s$ by taking action $a$, 
$R(s,a)$ provides the immediate reward of taking action $a$ in state $s$, and $\gamma$ is the discount factor that balances the importance of immediate rewards against future rewards. 
In an MDP, the future state depends solely on the current state and action, regardless of the previous history.
Solving MDPs involves finding an optimal policy that maximizes the expected averaged sum of rewards.
This policy can be deterministic or stochastic, and it can be represented by a function $\pi(s)$ that returns the action to be taken in state $s$.

In a reinforcement learning model, the policy $\pi(s)$ is implemented by an artificial Neural Network~(NN)~\cite{Zou2009}, 
which is trained through executing different episodes of agent-environment interaction.
At the end of each episode (or batch of episodes), the average episode reward is computed and used to update the weights of the NN using the backpropagation algorithm~\cite{Rojas1996}.

Many optimization methods have been developed in recent years to improve reinforcement learning convergence and stability during training~\cite{ozalp2020}. 
In our work, we have obtained the best results using the Proximal Policy Optimization~(PPO)~\cite{schulman2017proximal}. 
The idea behind PPO lies in restricting the policy updates to avoid too large changes in the policy, which may lead to instability during training and,
as people say in jargon, to ``fall off the cliff''. Namely, too large updates may lead the agent to regions with low expected cumulative reward, that
may take a very long time to escape from. This is implemented in practice by clipping the loss function to discourage big policy updates:
\begin{equation}
  \mathcal{L}_{t} (\theta) = \hat{E}_t [ \min ( r_t(\theta) \cdot \hat{A}_t, \rm{clip}(r_t(\theta), \epsilon) \cdot \hat{A}_t ) ]\;.
\end{equation}
The probability ratio $r_t(\theta)$ is defined as:
\begin{equation}
  r_t(\theta) = \frac{\pi^t_{\theta}(a_t | s_t)}{\pi^{t-1}_{\theta}(a_t | s_t)}\;.
\end{equation}
It evaluates the change in the policy going from step $t-1$ to $t$, and is limited in the final loss through the \emph{clip} function:
\begin{equation}
  \rm{clip}(r_t(\theta), \epsilon) = \begin{cases}
    1 + \epsilon      & \text{if } r_t(\theta) > 1 + \epsilon, \\
    1 - \epsilon      & \text{if } r_t(\theta) < 1 - \epsilon, \\
    r_t(\theta) & \rm{elsewhere}.
  \end{cases}
\end{equation}
This way big changes in the policy will still have a limited impact on the loss. This clipped ratio is then multiplied by
the advantage factor $\hat{A}_t$, as in the non-clipped case, which measures how good is on average the action taken at step t over
all the other possible actions.
Finally, the minimum between the clipped and non-clipped reward is taken to provide ``pessimistic predictions'' and, thus, discourage
larger jumps in the policy.
PPO often outperforms algorithms like Q-learning, in environments with continuous action spaces or partial observability, where Q-learning struggles due to the need for discretization. Compared to Soft Actor-Critic (SAC), which may achieve better asymptotic performance in highly stochastic settings due to its entropy regularization, we preferred PPO for its computational simplicity and lower variance in updates, while yielding satisfactory results.

We employed the \texttt{Stable\_Baselines3} library~\cite{JMLR:v22:20-1364} to define and train the algorithm. 
This library is built on top of \texttt{OpenAI Gym}~\cite{article_gym}, 
which provides a wide range of customizable environments for reinforcement learning tasks. 
Gradient optimization was performed with \texttt{PyTorch}~\cite{NEURIPS2019_9015}.

\section{Algorithm Implementation}\label{sec_methodology}

This section provides a detailed explanation of the proposed noise modeling algorithm, with a particular emphasis on the transformation of quantum circuits into input feature vectors suitable for neural network processing, 
as well as a comprehensive description of the policy, the training process and the datasets.

\subsection{Circuit Representation}\label{sec_circuit_representation}

To train the RL agent, we need to represent a quantum circuit as an array that can be readily processed by the policy neural network; we refer to this array as the Quantum Circuit Representation~(QCR). In the following, we refer to a circuit moment as a collection of gates that act on a non-overlapping set of qubits and are conceptually executed in parallel. The circuit depth is defined as the total number of circuit moments, assuming that the division of the circuit into moments is optimized to minimize their number.
The QCR has a shape of $[\,qubits,\: depth,\: encoding\,]$. The first dimension corresponds to the circuit's qubits, while the second dimension represents the circuit's moments.
The $encoding$ dimension encodes the information regarding gates and noise channels acting on a specific qubit at a specific circuit moment. 
Its dimension is determined by the total number of native gates and noise channels the model allows for.

In detail, considering a set of $n$ single-qubit native gates $\{G_i\}$, the initial $n$ entries of the $encoding$ will contain a {\it one-hot} encoding of our native set. Namely, in the presence of the gate $G_i$, all the entries are set to 0 except for the $i$-th entry that is set to 1. In the presence of a two-qubit gate, $CZ$ in our case, the entry relative to both the target and control qubit is set to 1.
Specifically, in our case, we use $n=2$ to encode the presence of an $R_x$ or $R_z$ gate.
The $n+1$ entry is set to $1$ if a two-qubit gate is present and $0$ otherwise, with the $CZ$ gate being the native two-qubit gate in this work. For single-qubit circuits, this entry is always zero.
The $n+2$ entry encodes the rotation angle of single-qubit gates, normalized to the range $[0,1]$.
The remaining entries detail the parameters of the noise channels in the following sequence: Depolarizing channel, Amplitude damping channel, $R_z$ coherent error, and $R_x$ coherent error. 
If a noise channel is absent, the corresponding parameter value is set to zero.
An example of the QCR for a two-qubit circuit, obtained using this procedure, is illustrated in Figure~\ref{fig_qcr}.

\begin{figure}
    \centering
      \begin{quantikz}[column sep=0.3cm]
        \lstick{$q_1$}& \gate{R_x(\pi)}\gategroup[2,steps=1,style={dashed,rounded corners, inner xsep=-0.5pt}]{Moment 1} &\ctrl{1} \gategroup[2,steps=2,style={dashed,rounded corners, inner xsep=-0.5pt}]{Moment 2} & \gate{\operatorname{Dep}(0.1)} & \gategroup[2,steps=2,style={dashed,rounded corners, inner xsep=-0.5pt}]{Moment 3} & &\\
        \lstick{$q_2$}& \gate{\operatorname{Coh_x}(0.1)}        &\ctrl{0}  & \gate{\operatorname{Dam}(0.2)}        & \gate{R_x(\pi/2)} & \gate{\operatorname{Coh_z}(0.05)} &
      \end{quantikz}

        \vspace{0.3cm}
        \setlength{\tabcolsep}{3pt}
        \renewcommand{\arraystretch}{1.15} 
        \begin{tabular}{@{}|ll|llllllll|@{}}
        \hline
        & & $R_z$ & $R_x$ & $CZ$ & $\theta$ & $\operatorname{Dep}$ & $\operatorname{Dam}$ & $\operatorname{Coh_z}$ & $\operatorname{Coh_x}$\\
        \hline
        \multirow{2}{*}{Moment 1}   & $q_1$ &0&1&0&0.5&0&0&0&0\\
                                    & $q_2$ &0&0&0&0&0&0&0&0.1\\
        \hline
        \multirow{2}{*}{Moment 2}   & $q_1$ &0&0&1&0&0.1&0&0&0\\
                                    & $q_2$ &0&0&1&0&0&0.2&0&0\\
        \hline
        \multirow{2}{*}{Moment 3}   & $q_1$ &0&0&0&0&0&0&0&0\\
                                    & $q_2$ &1&0&0&0.25&0&0&0.05&0\\
        \hline
        \end{tabular}
    
        \vspace{0.2cm}
        
        \caption{Example of a two-qubit quantum circuit (top) and its vector representation (bottom). A two-qubit circuit of depth three is represented as a tensor of size (2, 3, 8), where the first entry identifies the qubit, the second denotes the circuit moment, and the third specifies the type of gate or noise channel.}\label{fig_qcr}
\end{figure}

To enable the agent to adapt to circuits of varying depths, we further introduce the concept of kernel size $k$, similar to the kernels used in convolutional neural networks (CNN)~\cite{GU2018354}. 
The kernel size establishes a ``window'' that restricts the number of circuit moments the agent can observe at any given time. 
For instance, with $k=3$, the agent only observes the current moment and the immediately preceding and following ones. 
The window's center starts from the first moment and slides one position at each step until the circuit's end is reached. 
To ensure that the first and last moments of the circuit are at the center of a full-sized window, we apply a symmetric zero-padding of length $(k-1)/2$ at the edges of the QCR.
This approach is based on the heuristic assumption that a gate's noise is most influenced by its temporally proximate gates.
At a given moment $m$, therefore, a window $(m-\frac{k-1}{2},\,m+\frac{k-1}{2})$ is extracted from the complete QCR of the circuit, effectively yielding a fixed dimension $[\,qubits,\: k,\: encoding\,]$ tensor that is fed as input to the agent. 

\subsection{Policy}\label{sec_policy}
Agent's actions can be depicted following the same QCR schema. They are represented as matrices, whose individual rows and columns represent the qubits and the distinct noise channels respectively. Each entry of the action matrix is computed by a forward pass through the policy network, which consists of three main components. The Feature Extractor ($\operatorname{FE}$) takes as input the QCR ($x^{(QCR)}$) and maps it to a high dimensional latent feature space $x^{(feat)}$,
\begin{equation*}
    x^{(feat)} = \operatorname{FE}(x^{(QCR)}) \, .
\end{equation*}
To efficiently capture the correlations between different qubits and moments, we employ a CNN, which should be suited for processing two-dimensional data.
The Actor Policy ($\operatorname{A}_\pi$), is a simple Multi Layer Perceptron (MLP) which takes as input the latent features and computes the actual action the agent is going to take ($x^{(action)}$).
\begin{equation*}
    x^{(action)} = \operatorname{A}_\pi(x^{(feat)}) \, .
\end{equation*}
Moreover, as PPO is an actor-critic policy we have a separated critic policy $\operatorname{C}_\pi$ with the same MLP architecture but independent weights, whose role is to rate the action selected by the actor $x^{(rating)}$. 
The rating is extracted from the latent features,
\begin{equation*}
    x^{(rating)} = \operatorname{C}_\pi(x^{(feat)}) \, .
\end{equation*}
A schematization of the complete policy NN is reported in Figure~\ref{fig_policy}.
\begin{figure}[t]
\centering
    \begin{tikzpicture}[node distance=1.5cm]
    \node (obs) [block, align=center] {Observation ($x^{(QCR)}$)};
    \node (cnn) [block, below of=obs, align=center] {CNN features extractor ($\operatorname{FE}$)};
    \node (feat) [block, below of=cnn, align=center] {Features ($x^{(feat)}$)};
    \node (actor) [block, below of=feat, align=center, xshift=-2cm] {MLP actor ($\operatorname{A}_\pi$)};
    \node (critic) [block, below of=feat, align=center, xshift=2cm] {MLP critic ($\operatorname{C}_\pi$)};
    \node (rate) [block, below of=critic, align=center] {Rating ($x^{(rate)}$)};
    \node (action) [block, below of=actor, align=center] {Action ($x^{(action)}$)};
    \draw [arrow] (obs) -- (cnn);
    \draw [arrow] (cnn) -- (feat);
    \draw [arrow] (feat) -- (actor);
    \draw [arrow] (actor) -- (action);
    \draw [arrow] (feat) -- (critic);
    \draw [arrow] (critic) -- (rate);
    \end{tikzpicture}
    \vspace{0.2cm}
    \caption{Schematization of the policy neural network for the PPO algorithm. A common feature extractor ($\operatorname{FE}$), composed of a CNN, maps the QCR to a high-dimensional latent feature space. The resulting feature vector serves as input to both the actor policy ($\operatorname{A}\pi$) and the critic policy ($\operatorname{C}\pi$). The actor policy is responsible for selecting the best action for the agent, while the critic policy estimates the future reward for the action chosen by the actor.}
    \label{fig_policy}
\end{figure}

We have performed different tests to determine the optimal architecture and number of parameters for the NNs used in our policy. These hyperparameters slightly change with the number of qubits. Here, we report the best characteristics observed for circuits with one and three qubits. For the feature extractor, we used a single convolutional layer with 16 filters for single-qubit circuits and 32 filters for three-qubit circuits. This convolutional layer is followed by a dense layer with a ReLU activation function. The optimal number of output features for this dense layer is 64 for single-qubit circuits and 32 for three-qubit circuits.
Both the actor and critic policies are implemented as MLPs with a hidden dense layer containing 256 neurons. The total number of trainable parameters in the entire policy NN is on the order of $10^4$.
We experimented with increasing the number of parameters by adding convolutional layers and increasing the number of features in the feature extractor. However, we observed overfitting when the total number of parameters approached the order of $10^5$.

As detailed in Section~\ref{sec_noise}, in this work we consider a set of four possible noise channels: depolarizing, amplitude damping, and coherent errors $R_x$ and $R_z$. This means, that the output of our actor policy $\operatorname{A}_\pi$ is going to be a $[\,qubits,\: 4\,]$ tensor encoding the predicted parameters of the inserted noise channels at that specific moment. A value of zero corresponds to no noise channel of that type inserted. For example, in a two-qubit circuit, the action
\begin{equation*} 
\left[\begin{array}{cccc}
     0.1 & 0 & 0 & 0.2 \\ 
     0 & 0.05 & 0.3 & 0 
\end{array} \right]
\end{equation*}
indicates that a depolarizing channel with a depolarizing probability of $0.1$ and an $R_x$ coherent error with a rotation angle of $0.2$ are added to the first qubit. 
Simultaneously, an amplitude damping channel with a damping probability of $0.05$ and an $R_z$ coherent error with a rotation angle of $0.3$ are added to the second qubit.

A sensitive hyperparameter of the algorithm is the maximum allowed value for each noise channel's parameter, $P_{max}$, which controls in practice the span of the search space. A smaller $P_{max}$ yields a narrower parameter space making convergence faster. However, a too small value might limit the expressive capability of the model, precluding the agent the access to possibly large reward areas of the parameter space.
We obtained good results by setting $P_{max}$ as twice the effective depolarizing parameter derived from a RB experiment.

\subsection{Training procedure}\label{sec_training}

The agent is trained to add noise channels to a noiseless quantum circuit in order to reproduce the noise pattern observed when executing the circuit in the presence of some target noise (whether simulated or real when executing on the hardware).
This process is outlined in Figure~\ref{fig_training}.
\begin{figure}[ht]
\centering
    \begin{tikzpicture}[node distance=1.5cm]
    \node (action) [block, align=center] {\textbf{Action:} place error channels\\ with chosen parameter.};
    \node (policy) [block, right of=action, xshift=3cm, align=center] {\textbf{Policy:} convolutional\\ neural network.};
    \node (env) [block, below of=action, align=center] {\textbf{Environment:} quantum\\ circuit representation.};
    \node (rew) [block, below of=policy, align=center] {\textbf{Reward:} trace distance\\ between density matrices.};
    \draw [arrow] (policy) -- (action);
    \draw [arrow] (action) -- (env);
    \draw [arrow] (env) -- (rew);
    \draw [arrow] (rew) -- (policy);
    \end{tikzpicture}
    \vspace{0.2cm}
    \caption{Training process of the RL algorithm. At each step the RL agent receives an observation of the environment, the quantum circuit representation of a non-noisy circuit. The agent policy performs an action that consists in putting any number of noise channels at a given circuit moment. At the end of the circuit the reward is computed to minimize the trace distance of the density matrix of the original noisy circuit and the reconstructed one.}
    \label{fig_training}
\end{figure}

Each training episode begins with the agent receiving a randomly selected noiseless quantum circuit from the training set.
Then, for each of the circuit's moments, the agent observes the current QCR and takes an action: any combination of the selected set of noise channels, together with their corresponding noise parameters, is inserted in that precise moment, yielding an updated QCR.

The agent receives the reward at end of each episode and the policy's NN parameters are then updated to maximize future rewards.
The reward is taken to be a function of the distance between the target Density Matrix~(DM) ($\rho_{\rm true}$) and the DM of the noisy circuit generated by the agent ($\rho_{\rm agent}$) under some selected metric. We have tested different metrics such as a simple element-wise Mean Squared Error~(MSE), density matrix fidelity and Trace Distance~(TD)~\cite{metrics}. Trace distance proved to be the most suitable metric, as it is both easy to compute and phase-invariant. Moreover, as shown by some of the results presented in Section~\ref{sec_applications}, most of the information about the noise is contained in the diagonal elements of the density matrices \cite{HUANG2020100043}. Finally, this metric is specifically designed to quantify the experimental distinguishability between two quantum states.\\
TD can not be directly used as reward, so we have explored different functional forms including $1/\operatorname{TD}^n$ and $-\log(\operatorname{TD})$.
The most effective form, as determined through empirical testing, has been found to be
\begin{equation*}
    \operatorname{R}(\rho_{\rm agent}, \rho_{\rm true}) = \frac{1}{\alpha\operatorname{TD}(\rho_{\rm agent}, \rho_{\rm true})^2+\epsilon} \, ,
\end{equation*}
where $\epsilon$ is a small parameter introduced to prevent numerical instabilities and the hyperparameter $\alpha$ can be used to normalize the reward. This equation essentially penalizes high values of the TD with a low reward. The average episode reward, denoted as $\mathcal{R}$, is determined by averaging $\operatorname{R}$ over all the training circuits within a batch of episodes.
A summary of the training procedure is provided in Algorithm~\ref{alg:train}.
After numerous episodes, the agent is expected to learn the optimal placement of noise channels in a noise-free circuit to reconstruct the final density matrix of the real noisy circuit. 
Once trained, our algorithm should be capable of generalizing to previously unseen circuits, thereby enabling realistic noisy simulations.

\RestyleAlgo{ruled} 
\begin{algorithm}\label{alg:train}
\For{\texttt{episode} in \texttt{n\_episodes}}{
    \texttt{circuit} = random\_extraction(\texttt{training\_set})\;
    \For{\texttt{moment} in \texttt{circuit}}{
        \texttt{observation} = \texttt{agent}.make\_observation(\texttt{circuit}, \texttt{moment})\;
        \texttt{action} = \texttt{agent}.action(\texttt{observation})\;
        \texttt{circuit}.add\_noise(\texttt{action})\;}
    \texttt{generated\_dm} = extract\_density\_matrix(\texttt{circuit})\;
    \texttt{reward} = compute\_reward(\texttt{generated\_dm}, \texttt{ground\_truth\_dm})\;
    \texttt{agent}.update\_policy(\texttt{reward})\;
}
\caption{Agent training procedure.}
\end{algorithm}

\subsection{Dataset Generation}\label{sec_dataset}

The RL algorithm requires training, testing, and evaluation datasets~\footnote{Note that we use the convention more commonly used in the Physics field for naming the datasets, rather than the ML field one. In detail, we call \emph{testing} set the collection of data used to verify the progress over training, whereas we refer to \emph{evaluation} set for the data we used to present the final scores obtained by the model.}, which consist of ensembles of random quantum circuits and their corresponding final DMs ($\rho_{\rm true}$). 
These DMs serve as ground truth labels during the training phase of the algorithm (Section~\ref{sec_training}). 
In simulations, the $\rho_{\rm true}$ are computed analytically. However, for circuits executed on hardware, they can be obtained using quantum state tomography~\cite{PhysRevLett.109.120403} or more efficient techniques such as classical shadow state reconstruction~\cite{PRXQuantum.2.010307, Eisert_2020, PRXQuantum.2.030348}.

In this study, we utilize circuits with gates in the native gate set $\{\,R_x(\pi/2),\,R_z(\theta),\,CZ\,\}$ implemented in the quantum devices of the Technology Innovation Institute (TII)~\cite{qibolab_platforms_qrc}.
We also conducted preliminary tests using the native gates set of IBM quantum hardware~\cite{Santos_2016}, which employs the $CNOT$ gate as the two-qubit entangling native gate. 
Modifying the native gates in our algorithm is a simple process and has not led to any significant changes in performance.

The training set for single-qubit circuits is composed of circuits with a fixed number of Clifford gates extracted randomly. Clifford gates are chosen due to, both, their lower simulation cost and their large use in randomized benchmarking and shadow state estimation~\cite{2019npj, PhysRevA.77.012307, Eisert_2020, PRXQuantum.2.030348}. 
In our specific case, since we consider $R_x$ and $R_z$ gates, the only allowed rotation parameters are multiples of $\pi/2$.\\
For three-qubit circuits we have used a more sophisticated training set. 
Half of the training set is composed of circuits with a fixed depth with the gates and parameters extracted randomly. 
The second half of the training set is composed of Clifford circuits implementing randomly chosen three-qubit Clifford unitaries~\cite{bravyi_hadamard-free_2021}.
These circuits do not have a fixed number of gates or fixed depth. We observed an improvement in the generalization properties of the algorithm when using this mixed training set.

For the performance evaluation we have used two different datasets. The first dataset is composed of non-Clifford circuits with fixed depth. The results obtained with this set demonstrate that the algorithm, even if trained on Clifford circuits, maintains its ability to generalize.
The second dataset for performance evaluation, consist of Clifford circuits with varying depths. We have used this dataset to fit the RB noise model and compare it with the RL agent.

For the datasets used in simulations, we defined custom noise models. Specifically, we used two different custom noise models for circuits with one and three qubits. 
For single-qubit circuits, we applied a depolarizing channel with a depolarizing parameter of $\lambda=0.02$ after each $R_z$ gate, and an amplitude damping channel with decay parameter $\gamma=0.03$ after each $R_x$ gate.
A coherent $R_x(\theta_x)$ error with angle $\theta_x=0.04\cdot\theta$ is introduced after each $R_x(\theta)$ gate.
Similarly, a coherent $R_z(\theta_z)$ error is added after each $R_z(\theta)$ gate, with $\theta_z=0.02\cdot\theta$. 
This noise model is not intended to be realistic but to test our algorithm on a gate dependent noise model. 
For three-qubit circuits, we used a similar noise on rotation gates and added a depolarizing channel with a depolarizing parameter of $0.02$ and an amplitude damping channel with a decay probability of $0.03$ after each $CZ$ gate. 
In Section~\ref{sec_applications}, we have also used a lower error rate noise model to test the algorithm's ability to generalize to different noise models.
A comprehensive summary of the noise parameters used in the simulations is provided in Table~\ref{tab_noise_parameters}.

\begin{table}[ht]
\centering
\setlength{\tabcolsep}{6pt}
\begin{tabular}{|c|ccccc|}
\hline 
 Noise & Gates & $\lambda$ & $\gamma$ & $\theta_x$ & $\theta_z$ \\ 
\hline 
\multirow{2}{*}{1 qubit}    & $R_x(\theta)$ & 0 & 0.03 & 0.04$\cdot\theta$ & 0 \\
                            & $R_z(\theta)$ & 0.02 & 0 & 0 & 0.02$\cdot\theta$ \\
\hline
\multirow{3}{*}{\shortstack{3 qubits \\ (high noise)}}      & $R_x(\theta)$ & 0 & 0.03 & 0.04$\cdot\theta$ & 0 \\
                                                            & $R_z(\theta)$ & 0.02 & 0 & 0 & 0.03$\cdot\theta$ \\
                                                            & $CZ$ & 0.02 & 0.03 & 0 & 0 \\
\hline
\multirow{3}{*}{\shortstack{3 qubits \\ (low noise)}}       & $R_x(\theta)$ & 0 & 0.01 & 0.015$\cdot\theta$ & 0 \\
                                                            & $R_z(\theta)$ & 0.015 & 0 & 0 & 0.02$\cdot\theta$ \\
                                                            & $CZ$ & 0.015 & 0.01 & 0 & 0 \\
\hline
\end{tabular}
\caption{Noise models used in simulation to train and evaluate the RL algorithm in different conditions. The columns show the noise channels parameters: $\lambda$ for the depolarizing channel, $\gamma$ for the amplitude damping channel, $\theta_x$ and $\theta_z$ for the coherent error $R_x$ and $R_z$ respectively. The rows indicate the gates subject to the noise channel, a zero noise parameter means that the noise channel is not applied to that gate.}
\label{tab_noise_parameters} 
\end{table}

Generating the training set is straightforward for simulations where the DMs are computed analytically. However, performing full state tomography to obtain these values on quantum hardware can be time-intensive. For this reason, we conducted preliminary tests on one- and three-qubit simulated circuits to determine the minimal length of the training set.
We trained the algorithm on datasets of varying sizes (from $10$ to $10^3$), using the noise models introduced in Table~\ref{tab_noise_parameters}, and assessed the performance on a common evaluation set. We found that for three qubits, it is sufficient to train the algorithm on a dataset with more than $100$ circuits to avoid overfitting and achieve optimal performance. For single-qubit circuits, a dataset with about $20$ circuits is sufficient to obtain nearly optimal performance. This result may vary with the complexity of the noise model; however, it demonstrates that dataset generation is not a bottleneck for quantum chips with a few qubits.

\section{Results} \label{sec_results}

The following sections detail the results obtained from applying the proposed algorithm in both simulations and on quantum hardware.

\subsection{Simulations} \label{sec_simulation}

\begin{figure*}
    \centering
    \includegraphics[trim=0 0 25.3cm 0, clip, width=0.48\textwidth]{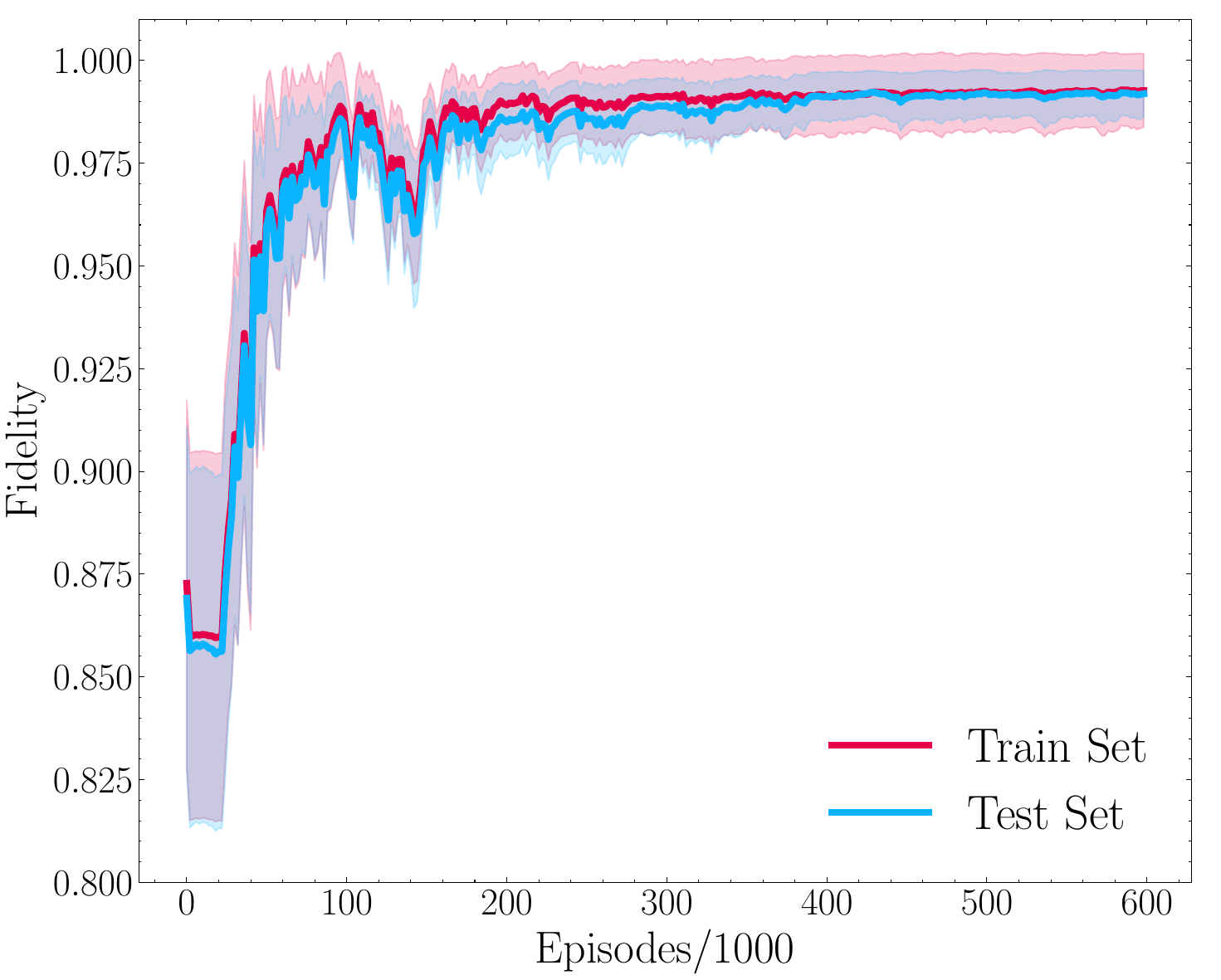}
    \includegraphics[trim=0 0 25cm 0, clip, width=0.48\textwidth]{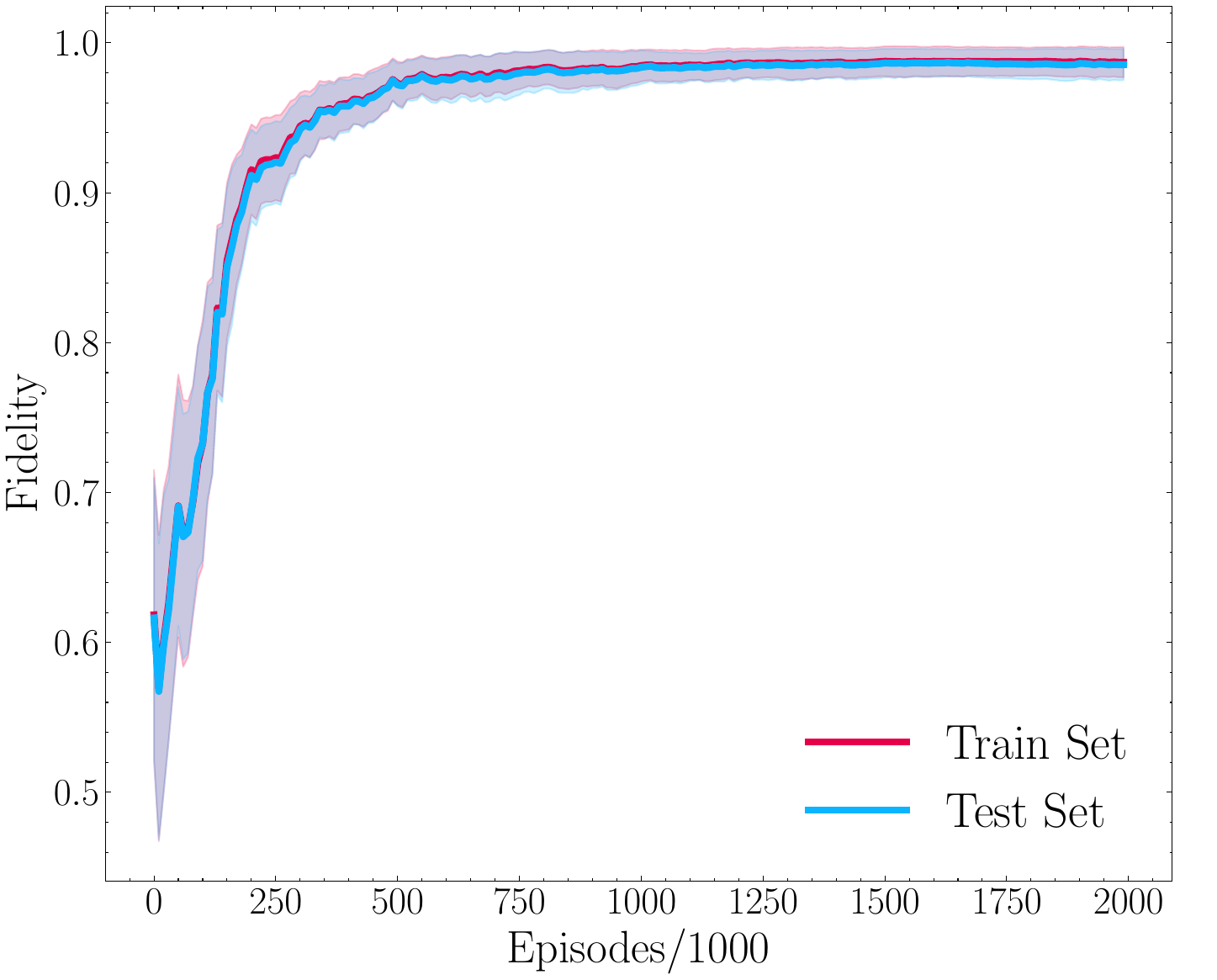}
    \caption{Average density matrix fidelity throughout the training process of the RL agent on single-qubit circuits (left) and three-qubit circuits (right). The metrics were evaluated on a dataset of 100 circuits for the single-qubit case and 800 circuits for the three-qubit case, using 80\% for the training set and 20\% for the test set. Shaded regions represent the standard deviation.\\
    In the single-qubit case, convergence of the training process was reached after approximately $4\times 10^5$ episodes, achieving an average fidelity of about $0.99$ on the test set. For the three-qubit case, convergence was reached after approximately $1.5\times 10^6$ episodes, achieving an average fidelity of about $0.98$ on the test set. No significant overfitting was observed in either case.}
 \label{fig_train}
    \vspace{0.3cm}
    \includegraphics[trim=0 0 25.3cm 0, clip, width=0.48\textwidth]{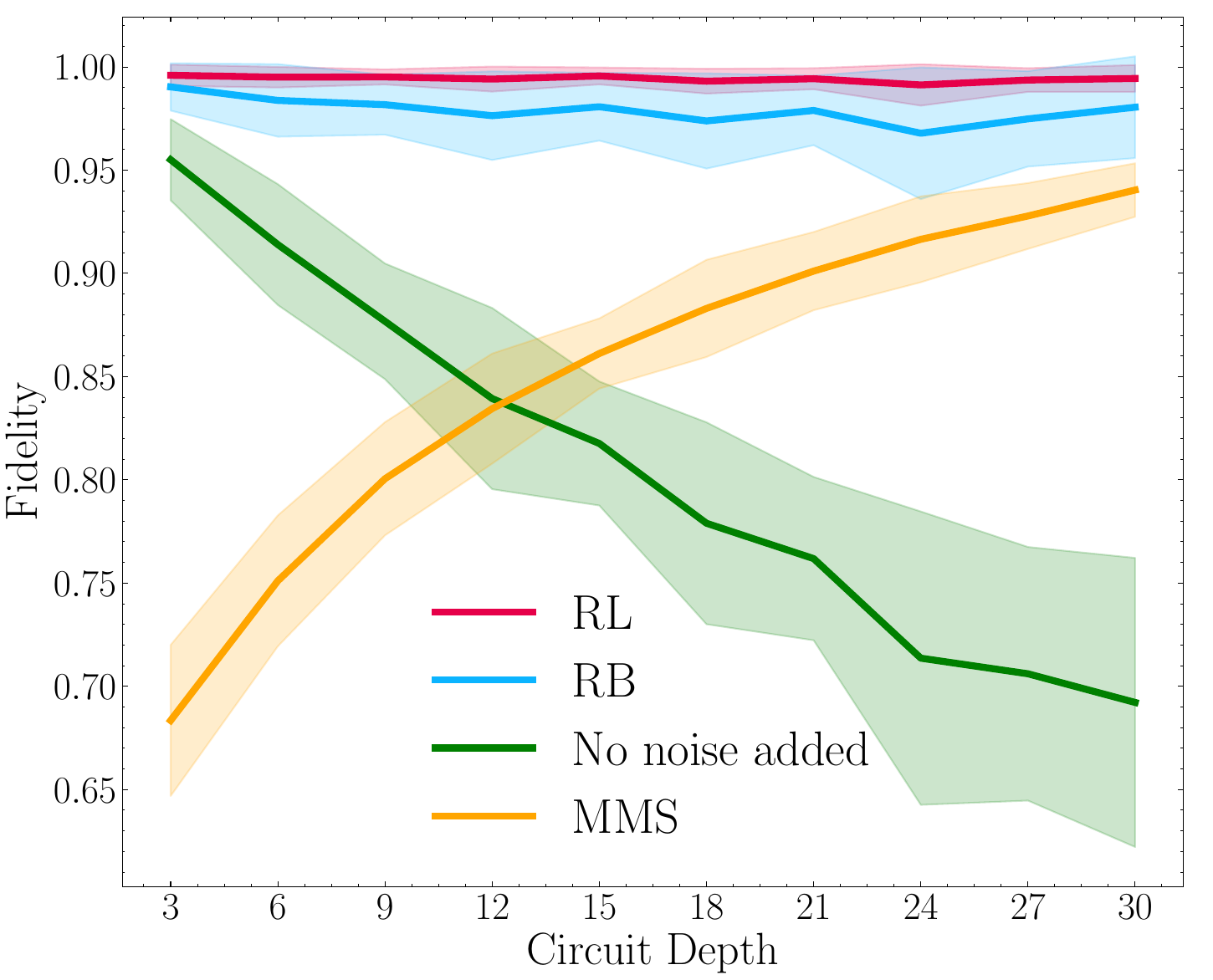}
    \includegraphics[trim=0 0 25.3cm 0, clip, width=0.48\textwidth]{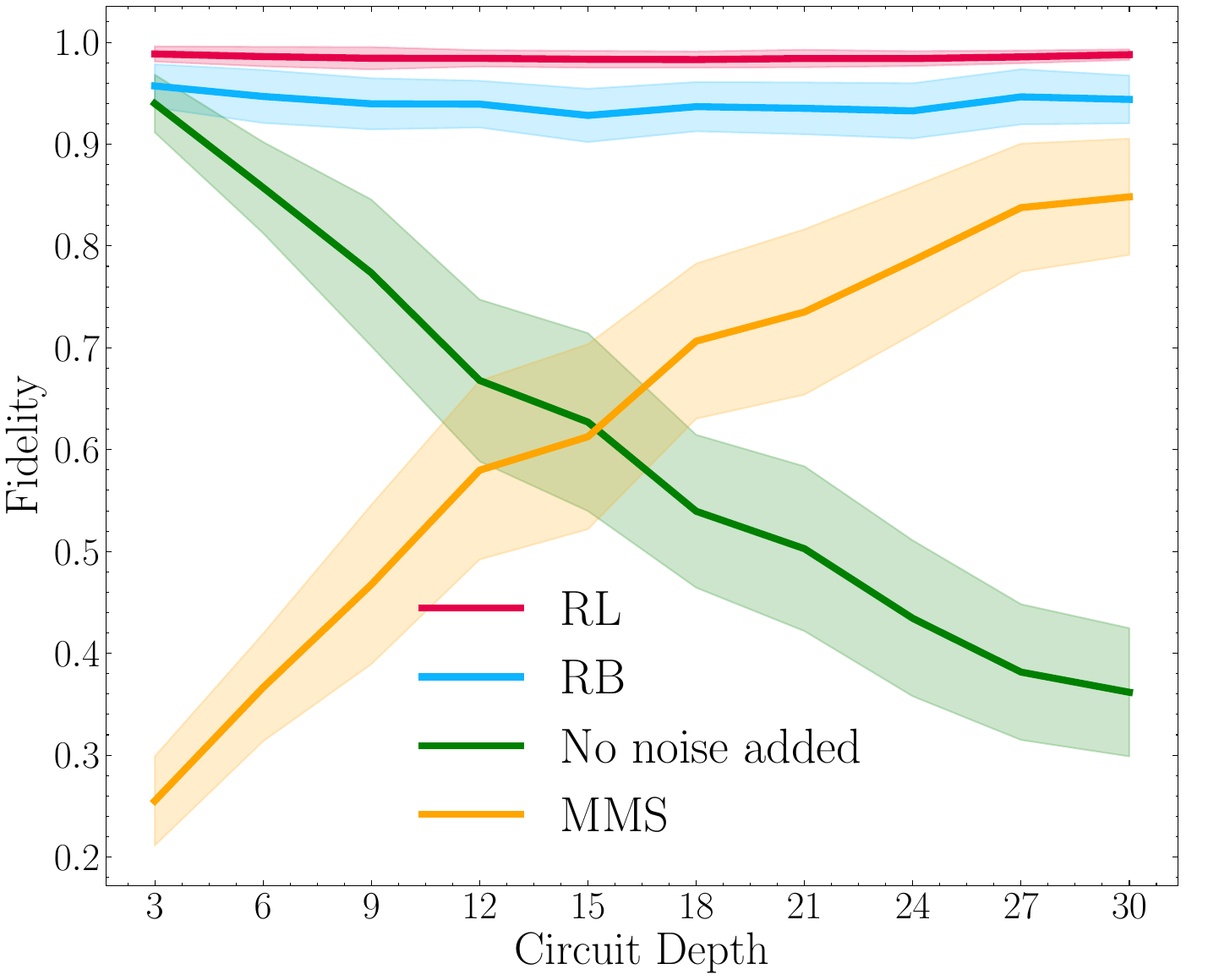}
    \caption{Performance evaluation of different noise models on Clifford circuits with depths varying from 3 to 30 for the single-qubit case (left) and the three-qubit case (right).
    Performance is measured using average density matrix fidelity; shaded regions represent the standard deviation. The RL agent was benchmarked against the RB method, the noiseless scenario, and the maximally mixed state (MMS). The absolute advantage of using the RL model is evident when compared to the case with no noise channel insertion. Moreover, in both cases, the RL model outperforms the RB model. Fidelity between the noisy circuit and the MMS is provided as a reference for the overall noise level; as the depth increases, the circuit’s behavior can approximate that of a completely depolarizing channel, converging toward the MMS.}
 \label{fig_rb}
\end{figure*}

Our study begins by training the RL agent to emulate the custom noise model, introduced in Section~\ref{sec_dataset}, on single-qubit circuits. To train the model, we generated a dataset of 100 random Clifford circuits of depth 10. We have used $80\%$ of the dataset for the training, reserving the remaining $20\%$ for the test set. 
The performance of the agent is evaluated by determining the average fidelity between the DMs it produces and the actual noisy DMs. These values are computed across all episodes for both the training and test sets, as shown in Figure~\ref{fig_train}.
The training process converges after roughly $4\times 10^5$ episodes, achieving an average fidelity of about $0.99$ on the test set. The RL agent effectively learns to simulate the noise, exhibiting no signs of overfitting.
For the evaluation set, we used a dataset of 100 random non-Clifford circuits with a depth of 15. The average fidelity of the RL agent on the evaluation set is $0.993$, with a standard deviation of $0.003$. This result shows that the agent is able to correctly generalize to non-Clifford circuits. In this kind of circuits, every possible rotation angle is allowed, while the training set consists of Clifford circuits with fixed rotation angles multiples of $\pi/2$.
A summary of the performance obtained by the model during the training and testing is reported in Table~\ref{tab_sim_results}.

To further assess the model's generalization capability, we evaluate it on random Clifford circuits of varying depths, from 3 to 30. Also in this case we evaluate the performance using the average fidelity between the DMs generated by the model and the actual noisy DMs. Figure~\ref{fig_rb} compares the performance of the RL agent with the RB noise model described in Section~\ref{sec_noise}). We also offer a comparison with two limit cases: the DMs of noiseless simulation and the DM of the maximally mixed state (MMS). The noiseless simulation serves as a baseline, representing an agent that does not apply any noise channels, and thus provides a reference for evaluating the benefit of using the RL agent. The maximally mixed state (MMS) is included as an additional reference to indicate the total noise level; as circuit depth increases, the system's behavior can approximate that of a completely depolarizing channel, converging toward the MMS.
The RL agent demonstrates its adaptability to circuits of different depths, consistently outperforming RB. 
This performance suggests that while RB categorizes all noise sources as depolarizing, our algorithm can discern the specific characteristics of the noise.\\

\begin{table*}[t]
\centering
\setlength{\tabcolsep}{8pt}
\begin{tabular}{|c|cccc|}
\hline
 & Training set & Test set & Evaluation set & Clifford depth 30 \\
\hline
1 qubit & $0.99\pm 0.01$ & $0.99\pm 0.01$ & $0.993\pm0.003$ & $0.99\pm 0.01$ \\
3 qubits & $0.98\pm 0.02$ & $0.98\pm 0.02$ & $0.98\pm 0.01$ & $0.98\pm 0.01$ \\
\hline
\end{tabular}
\caption{\label{tab_sim_results}. Summary of the RL model performance across different training, test, and evaluation sets for single- and three-qubit circuits. Performance is reported as the average fidelity between the reconstructed and original density matrices, with standard deviation as the error. The training and test sets consist of Clifford circuits for the single-qubit case, and both Clifford and non-Clifford circuits for the three-qubit case. The evaluation set includes non-Clifford circuits of depth 15 in both cases. Fidelity on Clifford circuits of depth 30 is also reported as an alternative evaluation set.}
\end{table*}

We extended our simulation to three-qubit circuits to evaluate performance in the presence of two-qubit gates. We have used the high noise model on three qubits reported in Table~\ref{tab_noise_parameters}.
The training set consists of 800 three-qubit circuits, divided into 80\% for training (640 circuits) and 20\% for testing (160 circuits). This training set is composed of both Clifford and non-Clifford circuits as described in Section~\ref{sec_dataset}.
The evolution of the average DMs fidelity throughout the training process is shown in Figure~\ref{fig_train}. 
While the algorithm is capable of learning the noise, the convergence is slower than in single-qubit circuits due to a larger action space requiring more episodes for exploration. 
Convergence is reached after approximately $1.5\times 10^6$ episodes achieving an average fidelity of about $0.98$. No signs of overfitting can be observed during the training phase.\\
For evaluation, we used a dataset of 200 random non-Clifford circuits of depth 15, where the RL agent achieved an average fidelity of $0.98$ with a standard deviation of $0.01$.

Mirroring our previous single-qubit circuit experiments, we evaluated the performance of our RL agent against both the RB method and the limit cases of noiseless circuits and MMS across a variety of circuit depths. This comparison is reported in Figure~\ref{fig_rb}. In these scenarios, the RL agent consistently demonstrates its adaptability to circuits of different depths, surpassing the performance of the RB method for all circuit lengths.\\
A summary of the RL model performance on training, test, and evaluation sets is reported in Table~\ref{tab_sim_results}.\\

\begin{figure}
    \centering
    \includegraphics[width=\linewidth]{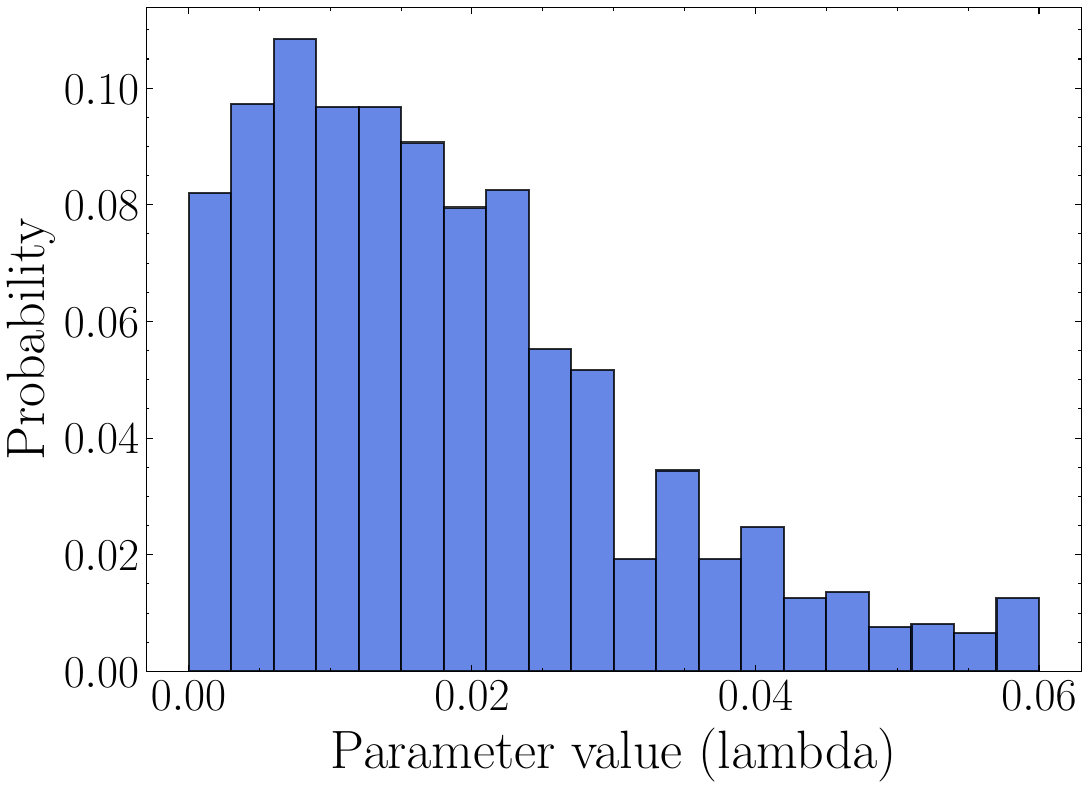}
    \includegraphics[width=\linewidth]{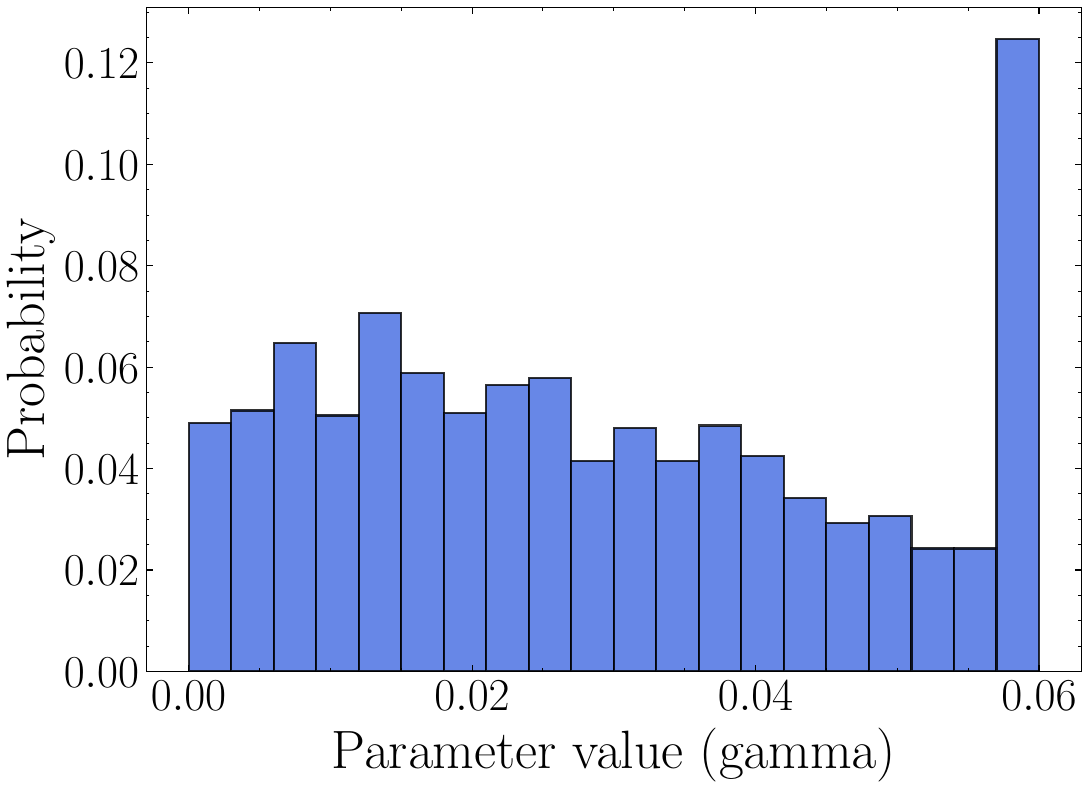}
    \caption{Distribution of the noise channels parameters, depolarizing parameter (lambda) top and 
    the amplitude damping parameter (gamma) bottom, inserted by the RL agent on a dataset of 100 random three-qubit Clifford circuits of depth 15. 
    The average values of lambda and gamma are respectively $0.018$ and $0.029$, which are very close to the real parameters of the noise model used in the simulation ($0.02$ and $0.03$ respectively).}\label{fig_noise_channels}
\end{figure}

In the results presented in this section we have shown that the RL agent is able to reproduce the noise model used in the simulation with high accuracy.
The limitation of this algorithm, which is a common limitation of ML algorithms, is that it is difficult to extrapolate information about the noise model.
Learning information about specific features of the noise may be useful to improve the reliability of the chip, for example by addressing the most problematic gates.
However, the RL agent behaves like a ``black box'' making it difficult to understand the reasons behind its decisions.
A deep understanding of the agent behavior would require the application of explainable ML techniques~\cite{xai1,xai2,xai3}, which is beyond the scope of this work.
In order to have a few insights about the noise model learned by the RL agent, we can keep track of the number of noise channels and parameter values inserted by the agent.
For this reason we have generated a dataset composed of 100 random three-qubit Clifford circuits of depth 15, and we have applied the high noise model reported in Table~\ref{tab_noise_parameters} to these circuits.
We have applied the RL algorithm to this dataset, and we have recorded the number of depolarizing and amplitude damping channels inserted by the agent and their parameters.
Figure~\ref{fig_noise_channels} shows the distribution of the parameters of these noise channels.
It is interesting to observe that, while the distributions are not peaked around the real parameters of the noise model, the average values of the depolarizing parameter (lambda) and 
the amplitude damping parameter (gamma) are respectively $0.018$ and $0.029$, which are very close to the real parameters of the noise model used in the simulation ($0.02$ and $0.03$ respectively).
The distributions of the parameters lambda and gamma are significantly different. The former is peaked around lower values and the shape is similar to a Poisson distribution, 
while the latter has a more uniform shape. Moreover, we can observe that the distribution of the amplitude damping parameter has a peak on the maximum value of $0.06$. This means 
that the agent is constrained by the maximum value of the parameter.
The average number per circuit of noise channels inserted by the agent is $19.7$ for the depolarizing channel and $20.2$ for the amplitude damping channel.
Also, this number is coherent with the noise model used in the simulation, where, for random generated circuits of depth 15, we expect to have about $20$ depolarizing and amplitude damping channels per circuit.
This study shows that we can extract some average noise parameters from the RL agent by looking at the number of inserted noise channels and their parameters.
However, this does not provide a complete understanding of the noise model learned by the agent and learning specific features of the noise.
The application of explainable ML techniques to this algorithm is a possible future direction of this work.

\subsection{Quantum hardware}\label{sec_hardware}
\begin{figure*}
    \centering
    \includegraphics[width=\linewidth]{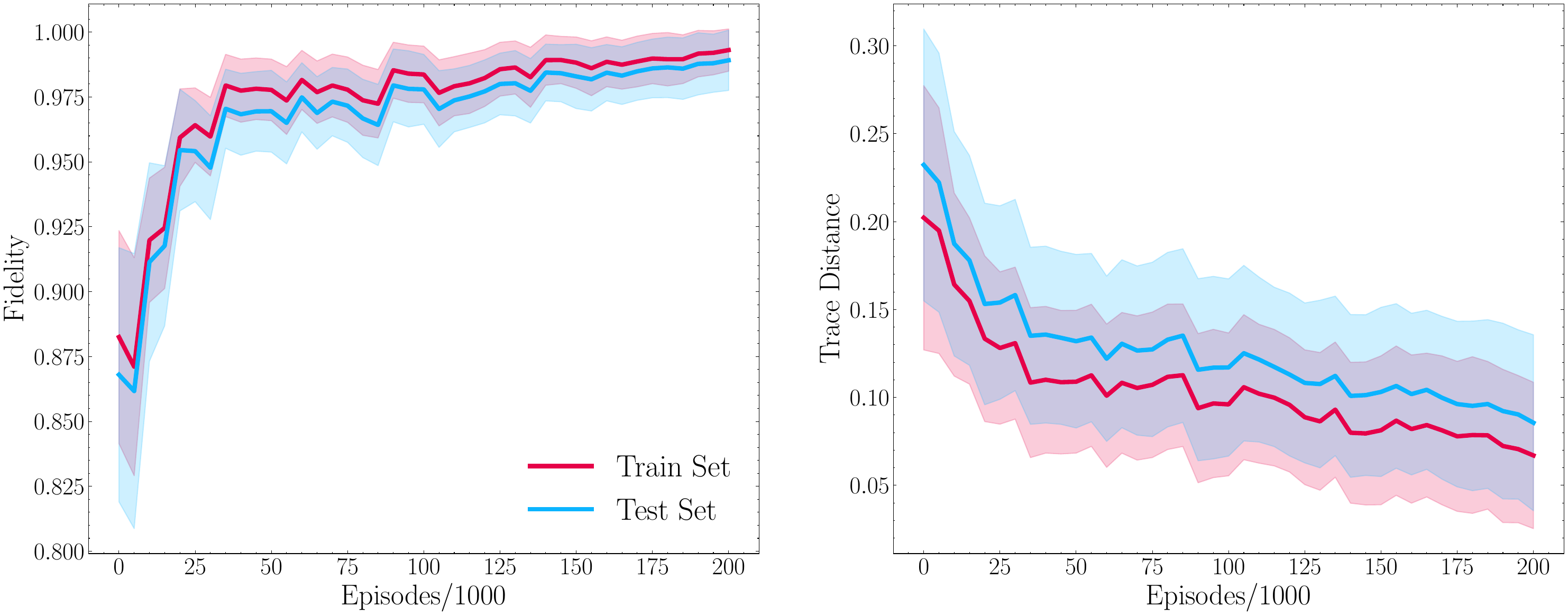}
    \caption{Average density matrix fidelity (left) and trace distance (right) throughout the training process of the RL agent on single-qubit circuits executed on a superconductive quantum chip. The metrics have been evaluated on a dataset of 60 circuits using 80\% for the training set and 20\% for the test set; shaded regions represent the standard deviation. Convergence was achieved after approximately $2 \times 10^5$ episodes, reaching an average fidelity of $0.99$ on the test set. No evident signs of overfitting were observed.}\label{fig_1q_hard_train}
\end{figure*}

To test our algorithm on real quantum hardware, we used a single-qubit in a superconducting transmon chip~\cite{doi:10.1126/science.1231930}. This 17 qubits chip has been produced by QuantWare~\footnote{\url{https://www.quantware.com}} and hosted at the Technology Innovation Institute of Abu Dhabi. The single-qubit gate fidelity, obtained with RB, is $0.996$ with a readout fidelity of about $0.96$. To compute the DMs of the circuits, we used state tomography, running $4\times10^3$ shots to compute each matrix. 
We attempted to mitigate measurement noise in advance to improve the fidelity of the DMs~\cite{nachman_unfolding_2020}. However, we observed that the RL agent performs better in learning the noise model without measurement error mitigation.

To train the algorithm, we collected a dataset consisting of 60 random circuits of depth 10, employing $80\%$ of these circuits for the training set and the remaining $20\%$ for the test set. Figure~\ref{fig_1q_hard_train} reports the average DM fidelity and TD during training for the first  $2\cdot10^5$ episodes, after this episode no performance improvements have been observed. At convergence, the model reached an average fidelity of $0.99$, with no evident signs of overfitting. This result is similar to the one obtained with simulations for single-qubit circuits.

Using a qubit with high gate fidelity makes the training process more challenging.
The ground truth density matrices are affected by both gate errors, which are learned by the RL algorithm, and by shot noise and measurement errors. In the high gate fidelity regime, measurement noise and shot noise can have an impact similar to that of gate noise. These errors in the density matrices make the reward signal less precise, thereby worsening the convergence of the training process.
For our dataset, we have computed the average error on the trace distance introduced by measurement errors, shot noise, and the additional gates needed to perform state tomography (see Appendix~\ref{app:dm_error}).
The obtained value is $0.036$, which explains the large error bars observed during training, as shown in Figure~\ref{fig_1q_hard_train}.
To mitigate this problem it would be useful to train the algorithm on longer circuits so that more gate errors can accumulate. In our case, to help the training process, it has been fundamental to set the maximum noise parameter value $P_{max}$ to a low value. As described in Section~\ref{sec_policy}, we have set $P_{max}$ to $0.008$, twice the decay parameter obtained with RB.\\

\begin{figure*}
    \centering
    \includegraphics[width=\linewidth]{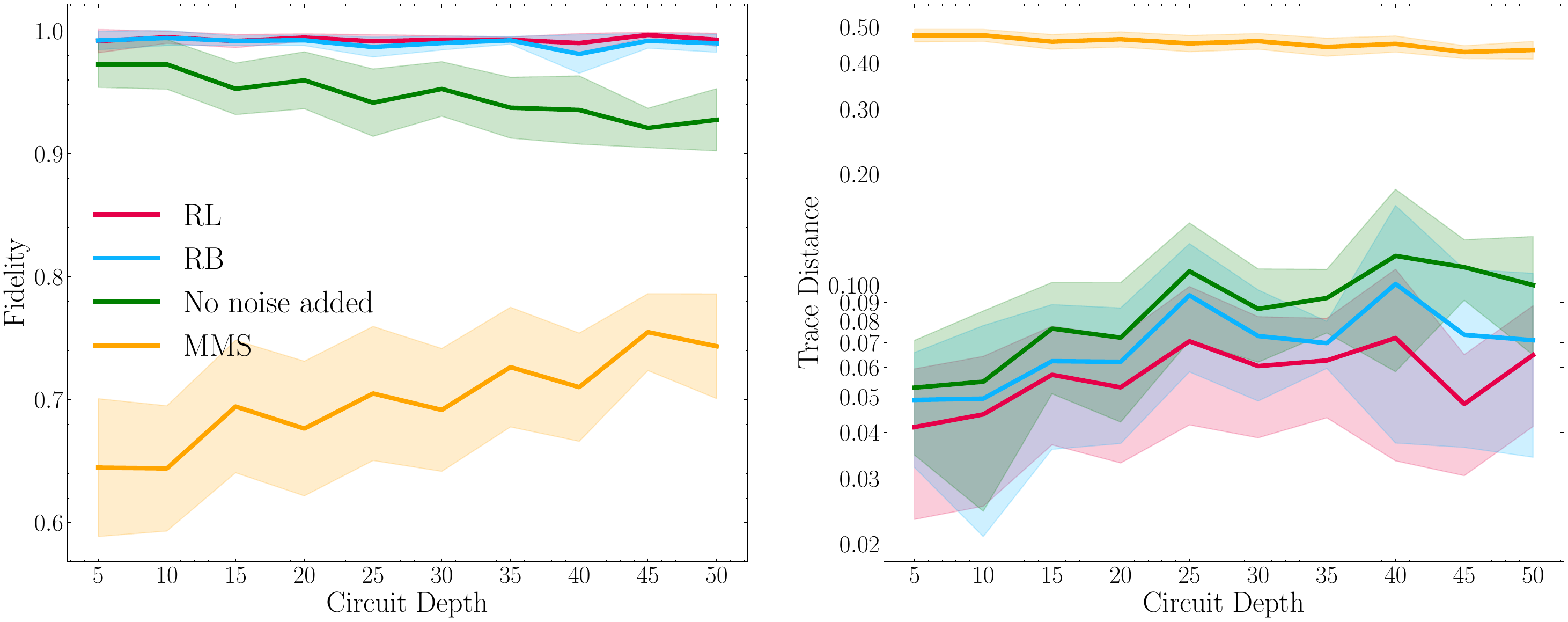}
    \caption{Performance evaluation of different noise models on single-qubit Clifford circuits with depths ranging from 5 to 50. The performance is assessed using average DM fidelity (left) and trace distance (right), with respect to the real noisy DMs obtained from execution on quantum hardware. The RL agent is benchmarked against the RB method, the noiseless scenario, and the maximally mixed state (MMS). In the trace distance plot, a logarithmic scale is used to enhance the distinguishability between RB and RL. The lower level of noise, compared to the simulation, can be inferred from the fidelity with the MMS, which does not reach the same values as in the simulation. A comparison between the RL and RB models shows that the former outperforms the latter on average; however, due to large errors, their performances are comparable within one standard deviation.}\label{fig_1q_hard_bench}
\end{figure*}

The performance benchmarking of the RL agent with respect to RB has been performed as described in Section~\ref{sec_simulation}. For this test, we used circuits of length spanning from 5 to 50, 10 circuits for each length. As the noise level of the qubit is quite low, it is necessary to use circuits with higher lengths to extract the decay parameter for the RB method.

Figure~\ref{fig_1q_hard_bench} summarizes the performance of the trained RL agent in reconstructing noisy density matrices, compared with the RB method. The comparison also includes results for circuits without noise and for the maximally mixed state. To highlight differences between the RL and RB noise models, we evaluated the TD using a logarithmic scale. On average, the RL agent outperforms the RB method across all circuit lengths, although their performances are statistically comparable within one standard deviation. The relatively large standard deviation observed is primarily due to the limited number of circuits in the test set and the presence of measurement errors, which introduce additional uncertainty in the ground truth DMs.

The results obtained in this section differ from those obtained in Section~\ref{sec_simulation}, where the RL agent clearly outperformed the RB method. However, it is important to highlight that the RL agent requires fewer hardware resources than the RB method. 
In the training of the RL agent, a dataset of $60$ circuits with a depth of $10$ has been sufficient (for a total number of $600$ circuit moments). 
The dataset generation process on the employed hardware took just a few minutes, at the cost of some classical resources needed for training the algorithm. 
Conversely, to estimate the RB parameter with comparable precision, it is necessary to run a larger number of circuits.
In our case, we needed to run a total of $110$ circuits with an average depth of $27.5$, for a total number of $3025$ circuit moments.
Performing the RB protocol required approximately five times more quantum hardware resources than generating the training set for the RL agent.\\
The overhead given by the RL training process is negligible because it is performed on classical hardware and does not require extensive resources.
The NN used for the algorithm has a total number of parameters of about $10^4$, which does not require a large amount of memory to be stored. 
The training process can be performed on a standard laptop in minutes without even requiring a GPU.
Scaling the algorithm to circuits with a higher number of qubits may be a challenge as 
a bigger action space and observation space would require a larger NN to be trained.
Addressing this challenge is a potential direction for future work. One possible solution is to divide the agent into multiple sub-agents, each responsible for a subset of the qubits.

It is interesting to note that calibration parameters, as well as other external factors that may affect the noise levels of a quantum chip, tend to vary over time. This implies that the RL agent should be retrained regularly on a new dataset to accurately capture the current noise conditions. This issue affects all noise modeling techniques. It is difficult to define a fixed time window during which the RL model, or any noise model, remains valid. This validity period depends on various factors, including the characteristics of the chip, refrigerator, electronics, and even the external conditions of the laboratory. Under stable conditions, the model can be considered valid until a new calibration routine is performed, which typically occurs on a daily basis.

\section{Test on quantum algorithms} \label{sec_applications}
In this section we assess our model's performance on the Quantum Fourier Transform~(QFT) and Grover's algorithm circuits under simulated noise. These tests provide valuable insights into the generalization capabilities of the model and serve as a stress test and benchmark for overall performance.

The QFT, a quantum counterpart of the classical Fast Fourier Transform, is a fundamental component of numerous quantum algorithms, including the renowned Shor's algorithm for factoring~\cite{Shor_1997}. The QFT acts on a quantum state $|x\rangle$ of $n$ qubits as
\begin{equation}
    \operatorname{QFT}\ket{x} = \frac{1}{\sqrt{2^n}} \sum_{k=0}^{2^n - 1} e^{2\pi ixk / 2^n} \ket{k} \, .
\end{equation}
Being a unitary transformation, the QFT can be implemented as a quantum circuit. For a three-qubit system ($n=3$), the QFT circuit is shown in Figure~\ref{qft_circ}.
\begin{figure}
\centering
    \begin{quantikz}[row sep=0.2cm, column sep=0.3cm]
        & \gate{H} & \gate{R_2}& \gate{R_3} & \qw & \qw & \qw & \swap{2} & \qw\\
        & \qw & \ctrl{-1} & \qw  & \gate{H} & \gate{R_2} & \qw & \qw & \qw\\
        & \qw & \qw & \ctrl{-2} & \qw & \ctrl{-1} &  \gate{H} & \targX{} & \qw\\
    \end{quantikz} 
\caption{QFT circuit for three qubits. In this circuit, $R_k=R_z(2\pi/2^k)$ and $H$ denotes the Hadamard gate. The last SWAP gate can be omitted as this reordering can be handled classically. This circuit implementation has to be transpiled into native gates before being passed to the RL agent. }\label{qft_circ}
\end{figure}
The layer of SWAP gates included to reorder the qubits is omitted in our implementation, as this reordering can be handled classically if the QFT is the final operation in an algorithm.

Grover's algorithm, another cornerstone of quantum computing, is renowned for its ability to search unsorted databases with quadratic speedup compared to classical algorithms~\cite{grover1996fast}. 
The algorithm operates on a superposition of quantum states, and its goal is to find a specific state $\ket{w}$ that satisfies a certain condition defined by an oracle function. 
The key component of Grover's algorithm is the Grover iterate, a unitary transformation that contain the information of the oracle. The Grover iterate is typically repeated $\mathcal{O}(\sqrt{N})$ times to maximize the probability of measuring $\ket{w}$, where $N$ is the dimension of the system.
We considered a two-qubit system and the target state $\ket{11}$. This configuration requires only one Grover iteration to find the target state. We utilized an ancillary qubit to construct the oracle, leading to the final circuit shown in Figure~\ref{grover_circ}.
\begin{figure}
\centering
\begin{quantikz}[row sep=0.2cm, column sep=0.3cm]
& \qw      & \gate{H} & \ctrl{2} & \gate{H} & \gate{X} & \ctrl{2} & \gate{X} & \gate{H} & \qw  \\
& \qw      & \gate{H} & \ctrl{1} & \gate{H} & \gate{X} & \ctrl{1} & \gate{X} & \gate{H} & \qw  \\
& \gate{X} & \gate{H} & \targ{}  &  \qw     &  \qw     & \targ{}  &   \qw    &     \qw  & \qw  \\
\end{quantikz}
\caption{Grover's search algorithm circuit, the target state is $\ket{11}$ and an ancillary qubit is required. In this configuration it is sufficient to interrogate the oracle only one time to obtain the correct result. The circuit has been transpiled into native gates before being used as input for the RL agent.}\label{grover_circ}
\end{figure}

We transpiled the circuits to utilize the native gates $R_z$, $R_x$, and $CZ$. The gate number after transpilation and other significant circuit parameters are detailed in Table~\ref{tab_gates}. A random non-Clifford circuit of the same type used for the training set is also included as a benchmark.
It should be noted that the circuits implementing the QFT and Grover algorithms are substantially different from those used in the training set.
The main differences include increased circuit depth, a lower fraction of two-qubit gates, and structural variations. Some gate patterns present in these circuits were likely never encountered by the agent during training, as they would be unlikely to arise through random circuit construction.
All these factors contribute to making the generalization test performed in this section particularly challenging.

\begin{table}[ht]
\centering
\setlength{\tabcolsep}{8pt}
\begin{tabular}{|c|ccc|}
\hline
Circuit & Total gates & $CZ$ gates & Depth \\
\hline
QFT & 23 & 6 & 15 \\
Grover & 40 & 7 & 25 \\
Random & 23 & 12 & 15 \\
\hline
\end{tabular}
\caption{\label{tab_gates} Number of gates and circuit depth of the transpiled circuits for QFT and Grovers's algorithm.
A random circuit from the training set is also included as a benchmark to highlight the differences including depth and a higher fraction of two-qubit gates.}
\end{table}

We have tested the algorithms using both the high and low noise models for three-qubit circuits detailed in Section~\ref{sec_simulation}. 
This approach served a dual purpose. Primarily, we aimed to test the algorithm's generalization capabilities with noise models exhibiting lower error rates. 
Secondly, the noise model from Section~\ref{sec_simulation} completely masked the final result for Grover's algorithm. 
By using a noise model with lower error rates, we ensured that a peak at the target state in the final result remained discernible.
We evaluated the performance of our RL agent against the RB noise model and the limit case where no noise is added to algorithm's circuits. A comparison with the maximally mixed state is also given to give an estimate of the total level of noise.
The fidelity between the reconstructed and original noisy DMs for the different noise models is detailed in Table~\ref{tab_results}. We also report the result obtained with the random circuit as a benchmark. In all instances, the RL agent obtained the best performance. 

\begin{table}[ht]
\centering
\setlength{\tabcolsep}{6pt}
\begin{tabular}{|cc|cccc|}
\hline
Circuit                 & Noise & RL   & RB   & No noise added & MMS \\
\hline
\multirow{2}{*}{QFT}    & High  & \textbf{0.99} & 0.97 & 0.59 & 0.70 \\
                        & Low   & \textbf{0.99} & \textbf{0.99} & 0.78 & 0.52 \\
\hline
\multirow{2}{*}{Grover} & High  & \textbf{0.98} & 0.95 & 0.40 & 0.83 \\
                        & Low   & \textbf{0.98} & 0.96 & 0.65 & 0.64 \\
\hline
\multirow{2}{*}{Random} & High  & \textbf{0.99} & 0.93 & 0.61 & 0.62 \\
                        & Low   & \textbf{0.99} & 0.98 & 0.77 & 0.51 \\
\hline
\end{tabular}
\caption{\label{tab_results}Fidelity between the density matrix reconstructed with a noise model (RL agent, RB, the limit case where no noise channels are added and the MMS) and the ground truth noisy one. 
The result is reported for QFT and Grover's algorithm circuits for both a high and low noise models. The result obtained with a non-Clifford random circuit is also reported for comparison.
The RL agent outperforms, or is comparable, with respect to the RB model in all the studied cases.}
\end{table}

\begin{figure}[ht]
    \raggedleft 
    \includegraphics[width=0.49\textwidth]{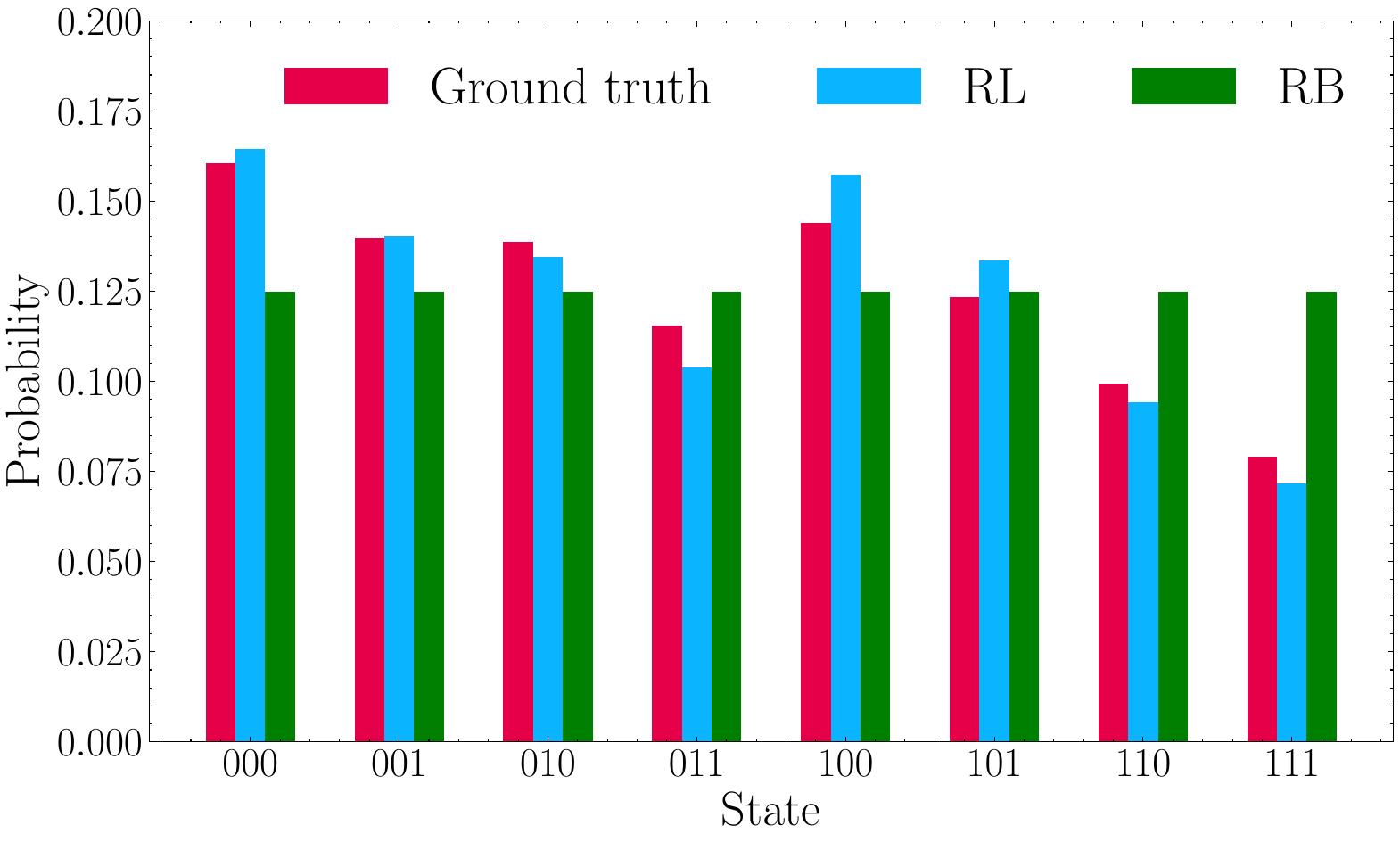}
    \includegraphics[width=0.49\textwidth]{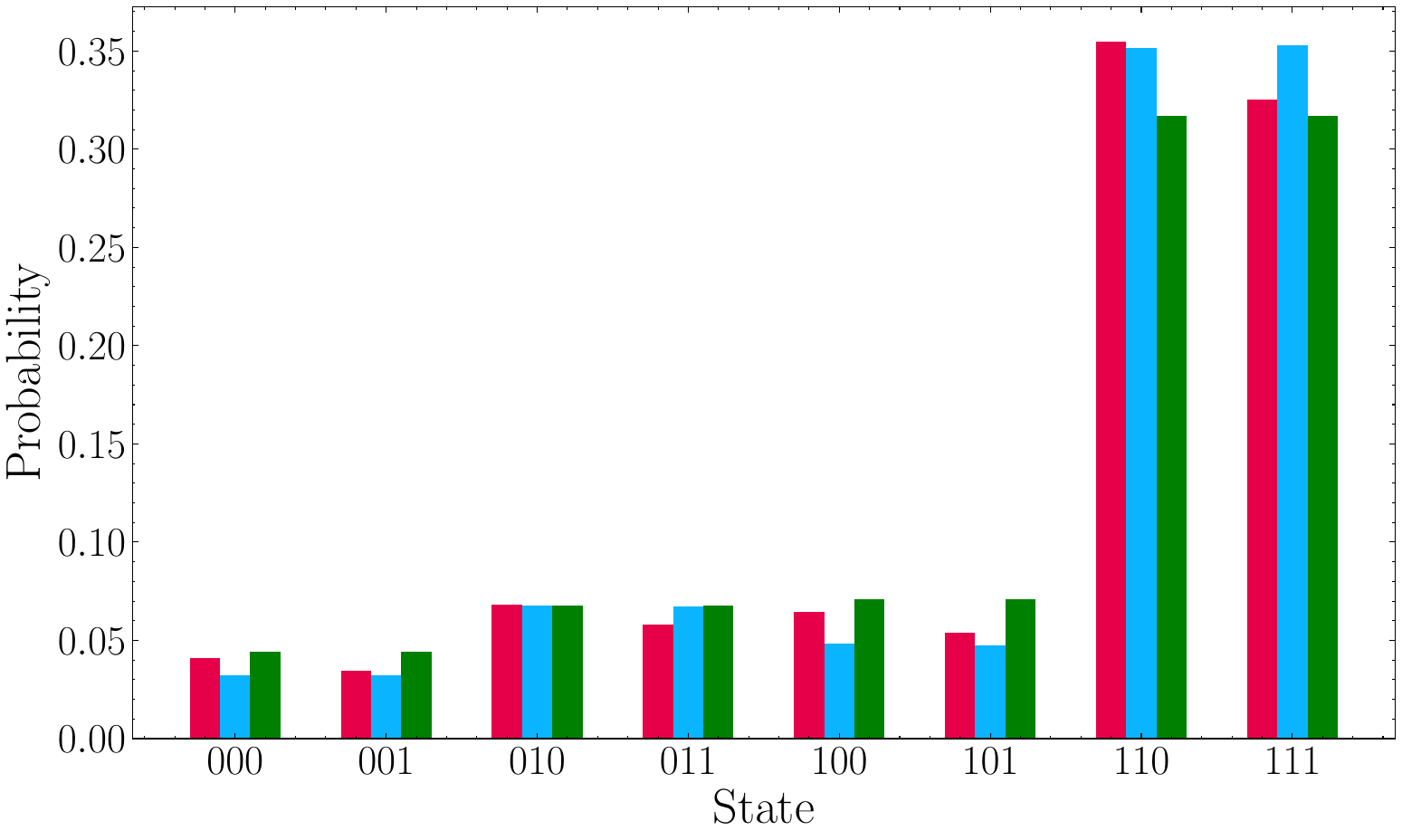}
    \includegraphics[width=0.49\textwidth]{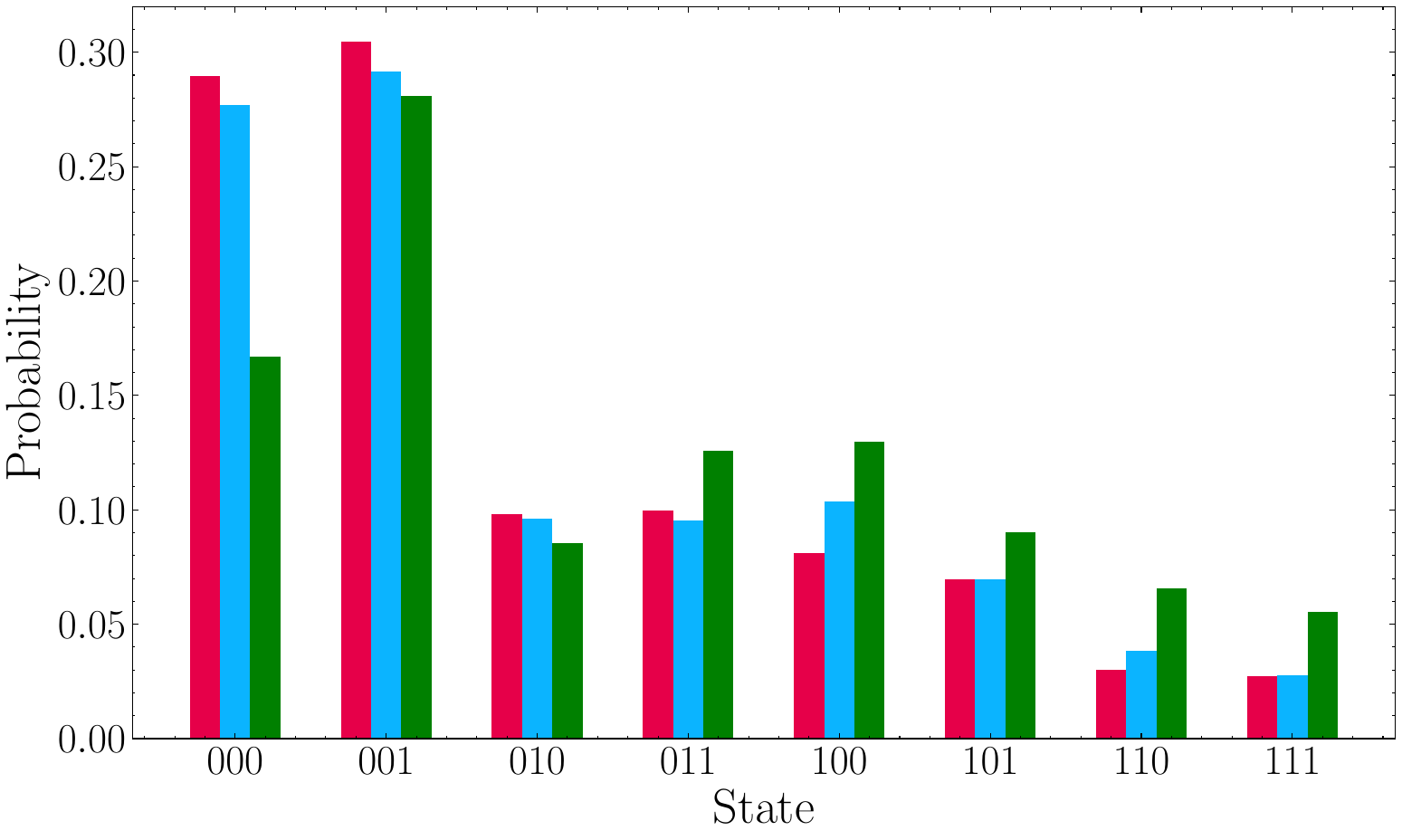}
    \caption{Computational basis states probabilities for the QFT circuit with a high error noise model (top), Grover's algorithm circuit with a low error noise model (mid) and a random non-Clifford circuit with a high error noise model (bottom). 
    The histograms show a comparison between probabilities obtained with the ground truth noise model, the RL agent noise model and the RB noise model. It is possible to observe that while the RB model tends to flatten the distribution, the RL agent is able to reconstruct the peaks with more fidelity.}
    \label{fig_algorithms}
\end{figure}

\begin{figure*}[t]
    \centering
    \includegraphics[width=0.85\textwidth]{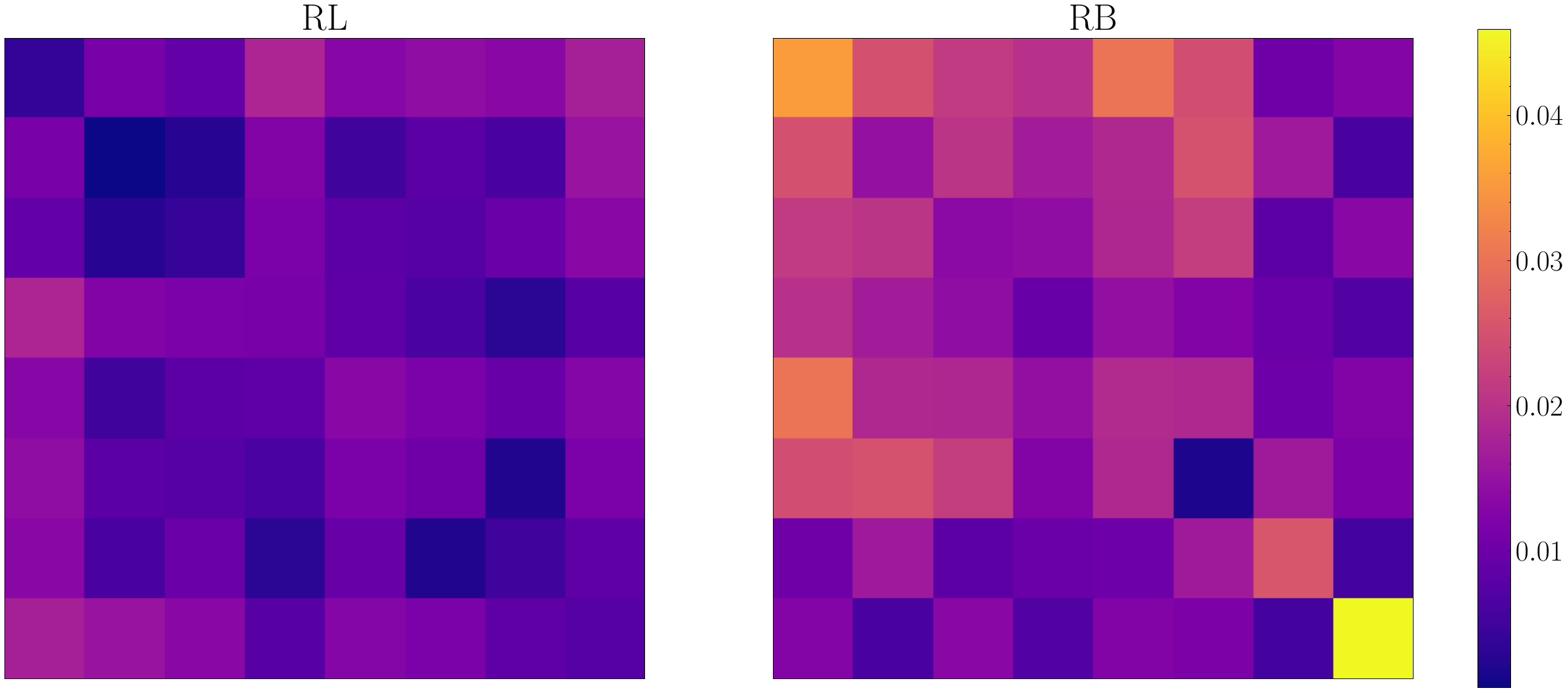}
    \caption{Heatmap of the absolute error between the ground truth noisy DM and the DM obtained with the RL agent (left) and RB model (right) for the three-qubit QFT circuit. It is possible to observe that the RL agent is able to better reconstruct the noisy DM not only in the diagonal part. This behavior has been observed in all the circuits tested in this section.}
    \label{fig_heatmap}
\end{figure*}

The final state probabilities for both the QFT and Grover's algorithm circuits, measured in the computational basis, are shown in Figure~\ref{fig_algorithms}. For the QFT circuit we report the result obtained with a high error noise model while for the Grover's circuit we report the result obtained with a low error noise model that does not destruct the expected result. Also in this case we report the result obtained with the random circuit previously introduced (high level noise model). The histograms compare the probabilities derived from the original noisy circuit, the reconstructed circuit using the RL agent, and the RB noise model. Given the depth of the circuits used, the RB noise model tends to average the output. In contrast, the RL agent, with a few exceptions, aligns more closely with the probabilities obtained from the original noise model. 
This alignment is particularly noticeable in the QFT simulation counts for the state $\ket{000}$. This state has the higher probability due to the noise model's amplitude damping channels, a feature that is successfully replicated by the RL agent but that is is not possible to replicate with the RB model. 

For a more detailed analysis, Figure~\ref{fig_heatmap} shows the absolute error between the DMs generated by the RL algorithm and the RB model compared to the ground truth noisy DM. The errors in the RL model are well distributed across the DM, while the RB model's errors tend to be higher along the diagonal. This effect was observed in many circuits, even during preliminary tests with the RL algorithm. As the depth of the circuit increases, the density matrix of noisy circuits tends to approach the MMS, resulting in most of the information being contained in the diagonal.

The results obtained in this section underscore the generalization capabilities of the proposed RL approach for noise modeling. The RL agent adapts to circuits with structures distinct from the random ones used in the training set and of significantly different depths. This allows the algorithm to be used for circuit simulations for interesting use cases like quantum algorithms and quantum machine learning when the exact noise model of the hardware chip is unknown.

\section{Conclusions}
This work presents a novel reinforcement learning algorithm for accurately replicating complex noise models in single- and multi-qubit quantum circuits. 
By learning noise patterns directly from experimental data, this approach significantly reduces heuristic assumptions about the underlying noise model, thereby enhancing generalization properties. 
Tests on simulated and quantum hardware circuits have demonstrated the model's ability to learn complex noise patterns and generalize to unseen circuits.
For a comprehensive evaluation and to show a possible use case, we tested the model on QFT and Grover's algorithm circuits. 
In all cases, the RL model consistently outperformed a common noise characterization method, randomized benchmarking both in the ability of reconstructing the density matrices 
and in the amount of quantum hardware resources needed. 

Possible future applications of the algorithm include not only reproducing the noise pattern of a specific hardware device. By learning the error patterns of qubits for specific gate types, the model could optimize the transpilation process~\cite{9259930}, thereby enhancing quantum algorithms fidelity. 
Furthermore, using the knowledge of the noise for its mitigation could be an interesting approach.
In this direction, it would be useful to employ explainable machine learning techniques to gain insights into the noise model learned by the RL agent.

The primary limitation of the current model is its scalability to quantum circuits with many qubits. This challenge manifests through two fundamental bottlenecks that must be addressed for practical deployment.
The first challenge concerns the growth of the action space. Scaling the model to larger qubit systems requires an increase in the number of possible actions, resulting in a larger expanded action space that complicates the training process.
The second challenge relates to the acquisition of ground truth density matrices via quantum state tomography, which requires an exponentially increasing number of measurements with respect to the number of qubits, making the approach intractable for large systems.
For the action space scalability issue, we propose partitioning large circuits into smaller, manageable subsystems, enabling parallel training of multiple specialized smaller models. Furthermore, replacing convolutional neural networks with graph neural networks could prove beneficial, as they can naturally encode qubit connectivity information and circuit topology into the model architecture.
To address the tomography bottleneck, one approach involves training the model directly on probability distributions derived from computational basis measurements, rather than full density matrices. 
Additionally, advanced machine learning techniques can be employed to reduce the measurement overhead required for quantum state tomography~\cite{Innan_2024, Schmale_2022}.\\

While these ideas require further validation, this work demonstrates that machine learning's application to learn noise patterns within small quantum circuits is a promising proof of concept that could lead to future advancements.

\vspace{0.5cm}
\section*{Declarations}
\noindent
\textbf{Author contribution:} 
S. Bordoni, A. Papaluca, P. Buttarini, A. Sopena: Methodology, Software, Writing.\\
S. Giagu, S. Carrazza: Project Administration, Review, Funding.\\

\noindent
\textbf{Funding:} 
This work is partially supported by the Technology Innovation Institute (TII), Abu Dhabi, UAE.\\
S.~B. and S.~G. are partially supported by ICSC - Centro Nazionale di Ricerca in High Performance Computing, 
Big Data and Quantum Computing, funded by European Union - NextGenerationEU.\\
A.~S. was supported through the Spanish Ministry of Science and Innovation grant SEV-2016-0597-19-4, 
the Spanish MINECO grant PID2021- 127726NB-I00, the Centro de Excelencia Severo Ochoa Program SEV-2016-0597 
and the CSIC Research Platform on Quantum Technologies PTI-001.\\

\noindent
\textbf{Data Availability:} 
Both the datasets and the code used for this study are available on GitHub: \url{https://github.com/qiboteam/rl-noisemodel}.\\

\noindent
\textbf{Open access:}
This article is licensed under a Creative Commons Attribution 4.0 International License, which permits use, sharing, adaptation, distribution and reproduction in any medium or format, as
long as you give appropriate credit to the original author(s) and the source, provide a link to the Creative Commons license, 
and indicate if changes were made. The images or other third party material in this article are included in the article's 
Creative Commons license, unless indicated otherwise in a credit line to the material. 

\bibliographystyle{apsrev4-2}
\bibliography{references}

\clearpage

\setcounter{equation}{0}\renewcommand\theequation{A\arabic{equation}}

\appendix

\section{Randomized benchmarking as a noise model} \label{app:rb}

The decay parameter $f$, extracted through randomized benchmarking (RB), can be used to construct a simple noise model for the quantum device. 
Consider one of the RB circuits consisting of $m$ gates. Let $\rho$ denote the density matrix of the initial pure state, and assume that each gate introduces a depolarizing channel characterized by a parameter $\lambda\in\qty(0,1)$.
The action of the overall noise channel $\mathcal{E}$ on the state after $m$ gates is
\begin{align}
    \mathcal{E}(\rho) = (1-\lambda)^m\rho+\qty(1-(1-\lambda)^m)\frac{I}{d}
\label{eq:depolarising_1}
\end{align}
where $d$ is the dimension of the system.
The fidelity between the noisy state $\mathcal{E}(\rho)$ and the ideal state $\rho$ is given by
\begin{align}
    F(\mathcal{E}(\rho),\rho) = (1-\lambda)^m + \frac{1-(1-\lambda)^m}{d} \, .
\label{eq:depolarising_2}
\end{align}
In our RB experiment, the initial state is $\ket{0}$, so the fidelity between the noisy state and the ideal state represents the probability of measuring the bit $0$ when measuring the final state in the computational basis.
Comparing~\eqref{eq:rb_decay} and~\eqref{eq:depolarising_2}, we can see that the depolarizing noise parameter $\lambda$ can be expressed in terms of the decay parameter $f$ as $\lambda=1-f$.

\section{Uncertainty estimation for the trace distance} \label{app:dm_error}

We estimate the error in the trace distance $T(\rho,\sigma)$ between the noisy density matrix $\rho$, obtained via state tomography, and the model-generated density matrix $\sigma$, by propagating the uncertainties of each element of $\rho$ as
\begin{equation}
    \Delta T(\rho,\sigma) = \sum_{i,j} \left| \frac{\partial T}{\partial \rho_{ij}} \right| \Delta \rho_{ij} \, ,
\end{equation}
where $\Delta \rho_{ij}$ is the uncertainty in the matrix element $\rho_{ij}$. The partial derivatives are approximated using finite differences.

We consider three main sources of error: (i) statistical fluctuations due to the finite number of shots, (ii) measurement errors, and (iii) errors introduced by additional gates required for basis changes in state tomography. 
Measurement errors are modeled as single-qubit bit-flip channels with probability $1-f_r$, where $f_r$ is the readout fidelity. 
Errors from basis-change gates are modeled as depolarizing noise with parameter $1-f$, as described in the main text.

A single-qubit density matrix can be written as
\begin{equation}
    \rho = \frac{1}{2}(I + \vec{r}\cdot\vec{\sigma}) \, ,
\end{equation}
where $\vec{r} = (r_x,r_y,r_z)$ is the Bloch vector and $\vec{\sigma} = (\sigma_x,\sigma_y,\sigma_z)$ are the Pauli matrices. The components of the Bloch vector are given by
\begin{equation}
    r_j = \text{Tr}(\sigma_j\rho) \, , \quad \text{for } j=x,y,z \, .
\end{equation}
Assuming $\rho$ is a noisy density matrix, the uncertainty on the Bloch vector components can be expressed as
\begin{equation}
    \Delta r_j = \sqrt{\frac{1-a_j^2}{N_s}} + (1-f)\abs{a_j} + \abs{a_j-\text{Tr}(\sigma_z\tilde{\rho}_j)} \, ,
\end{equation}
where $N_s$ is the number of shots, $a_j$ is the expectation value of $\sigma_j$ for the noiseless density matrix, and $\tilde{\rho}_j$ is the density matrix in basis $j$ including readout noise.
The uncertainty on the density matrix elements is then given by
\begin{equation}
    \Delta \rho_{ij} = \frac{1}{2}\vec{\Delta r}\cdot\vec{\sigma} \, , 
\end{equation}
where $\Delta\vec{r} = (\Delta r_x,\Delta r_y,\Delta r_z)$ is the vector of uncertainties on the Bloch vector components.

\end{document}